\documentclass[apj]{emulateapj}

\usepackage{apjfonts}
\usepackage{verbatim}
\usepackage{float}
\usepackage[export]{adjustbox}

\newcommand{\methanol}{CH$_3$OH}
\newcommand{\ammonia}{NH$_3$}

\newcommand{\HII}{H\,{\sc ii}}
 
\newcommand{\mjb}{mJy~beam$^{-1}$}

\newcommand{\mjyb}{mJy~beam$^{-1}$}

\newcommand{\kms}{km~s$^{-1}$}  
\newcommand{\mum}{$\mu$m}

\begin{document}

\title{VLA Survey of Dense Gas in Extended Green Objects: Prevalence of 25 GHz Methanol Masers}

\author{ A. P. M. Towner\altaffilmark{1*,2}, C. L. Brogan\altaffilmark{1}, T. R.  Hunter\altaffilmark{1},
C. J. Cyganowski\altaffilmark{3}, B. A. McGuire\altaffilmark{1$\star$}, R. Indebetouw\altaffilmark{1,2}, R. K. Friesen\altaffilmark{4},  C. J. Chandler\altaffilmark{5}}

\altaffiltext{1}{National Radio Astronomy Observatory, 520 Edgemont Rd, Charlottesville, VA 22903, USA}
\altaffiltext{2}{Department of Astronomy, University of Virginia, P.O. Box 3818, Charlottesville, VA 22903, USA}
\altaffiltext{3}{Scottish Universities Physics Alliance (SUPA), School of Physics and Astronomy, University of St. Andrews, North Haugh, St Andrews, Fife KY16 9SS, UK}
\altaffiltext{4}{Dunlap Institute for Astronomy and Astrophysics, University of Toronto, 50 St George St., Toronto, ON, M5S 3H4, Canada}
\altaffiltext{5}{National Radio Astronomy Observatory, 1003 Lopezville
  Rd, Socorro, NM 87801, USA}
\altaffiltext{*}{A.P.M.T. is a student at the National Radio Astronomy Observatory}
\altaffiltext{$\star$}{B.A.M. is a Jansky Fellow of the National Radio Astronomy Observatory}

\begin{abstract}
We present $\sim1-4''$ resolution Very Large Array (VLA) observations of four \methanol\/ $J_2-J_1$-$E$ 25~GHz transitions ($J$=3, 5, 8, 10) along with 1.3~cm continuum toward 20 regions of active massive star formation containing Extended Green Objects (EGOs), 14 of which we have previously studied with the VLA in the Class~I 44~GHz and Class~II 6.7~GHz maser lines \citep{Cyganowski2009}.  Sixteen regions are detected in at least one 25~GHz line ($J$=5), with 13 of 16 exhibiting maser emission. In total, we report 34 new sites of \methanol\/ maser emission and ten new sites of thermal \methanol\/ emission, significantly increasing the number of 25~GHz Class I \methanol\/ masers observed at high angular resolution.  We identify probable or likely maser counterparts at 44~GHz for all 15 of the 25~GHz masers for which we have complementary data, providing further evidence that these masers trace similar physical conditions despite uncorrelated flux densities.  The sites of thermal and maser emission of \methanol\/ are both predominantly associated with the 4.5 \mum\/ emission from the EGO, and the presence of thermal \methanol\/ emission is accompanied by 1.3~cm continuum emission in 9 out of 10 cases. Of the 19 regions that exhibit 1.3~cm continuum emission, it is associated with the EGO in 16 cases (out of a total of 20 sites), 13 of which are new detections at 1.3~cm.  Twelve of the 1.3~cm continuum sources are associated with 6.7~GHz maser emission and likely trace deeply-embedded massive protostars. 
\end{abstract} 


\section{Introduction}
\begin{deluxetable*}{lcccccccc}
\tablecaption{EGO Source Properties} 
\tablecolumns{9} 
\tabletypesize{\footnotesize} 
\tablehead{ \colhead{Source} & \colhead{$V_{\rm LSR}^{\rm a}$} & \colhead{Distance$^{\rm b}$} & \colhead{EGO$^{\rm c}$} & \colhead{IRDC$^{\rm d}$} & \colhead{H2O$^{\rm e}$} & \multicolumn{3}{c}{CH$_3$OH Masers (GHz)$^{\rm f}$}\\
              & \colhead{(\kms\/)}            & \colhead{(kpc)}             & \colhead{Cat} &      & \colhead{Maser} & \colhead{6.7$^{\rm g}$} & \colhead{44$^{\rm h}$} & \colhead{95$^{\rm i}$}
}
\startdata 
G08.67$-$0.35 & 35 & 4.1                                  & C13 & N  &  Y  & Y  & Y  & Y\\
G10.29$-$0.13 & 14 & 1.9                                    & 2 & Y  &  Y  & Y  & Y  & Y\\ 
G10.34$-$0.14 & 12 & 1.6                                    & 2 & Y  &  Y  & Y  & Y  & Y\\ 
G11.92$-$0.61 & 36 & 3.38$^{+0.33}_{-0.27}$ (3.5)           & 1 & Y  &  Y  & Y  & Y  & Y\\ 
G12.68$-$0.18 & 55 & 2.40$^{+0.17}_{-0.15}$ (4.4)           & 4 & Y  &  Y  & Y  & ?  & Y\\ 
G12.91$-$0.03 & 57 & 4.5                                    & 1 & Y  &  Y  & Y  & ?  & Y\\ 
G14.33$-$0.64 & 23 & 1.13$^{+0.14}_{-0.11}$ (2.3)           & 1 & Y  &  Y  & ?  & Y  & Y\\ 
G14.63$-$0.58 & 19 & 1.83$^{+0.08}_{-0.07}$ (1.9)           & 1 & Y  &  Y  & Y  & ?  & Y\\ 
G16.59$-$0.05 & 60 & 3.58$^{+0.32}_{-0.27}$ (4.2)           & 2 & N  &  Y  & Y  & ?  & Y\\ 
G18.67$+$0.03 & 80 & 4.8                                    & 1 & N  &  Y  & Y  & Y  & Y\\
G18.89$-$0.47 & 66 & 4.2                                    & 1 & Y  &  Y  & Y  & Y  & Y\\ 
G19.36$-$0.03 & 27 & 2.2                                    & 2 & Y  &  N  & Y  & Y  & Y\\ 
G22.04$+$0.22 & 51 & 3.4                                    & 1 & Y  &  Y  & Y  & Y  & Y\\ 
G24.94$+$0.07 & 42 & 2.8                                    & 1 & N  &  Y  & Y  & Y  & Y\\ 
G25.27$-$0.43 & 60 & 3.6                                    & 1 & Y  &  Y  & Y  & Y  & Y\\ 
G28.28$-$0.36 & 49 & 3.0                                    & 2 & Y  &  N  & Y  & N  & N\\ 
G28.83$-$0.25 & 87 & 4.8                                    & 1 & Y  &  Y  & Y  & Y  & ?\\ 
G35.03$+$0.35 & 53 & 2.32$^{+0.24}_{-0.20}$ (3.2)           & 1 & Y  &  Y  & Y  & Y  & Y\\ 
G45.47$+$0.05 & 61 & 8.40$^{+1.4}_{-1.1}$ (7.1)             & 1 & Y  &  Y  & N  & N  & N\\ 
G49.27$-$0.34 & 68 & 5.4                                    & 1 & Y  &  Y  & N  & Y  & Y
\enddata 
\tablenotetext{a}{LSRK velocities are the single dish NH$_3$ (1,1) values from \citet{Cyganowski2013}, except G08.67$-$0.35 which is the CH$_3$CN value from \citep{Purcell2006}. }
\tablenotetext{b}{Distances without errors are estimated from the LSRK
  velocity and the Galactic rotation curve parameters from \citet[][]{Reid2014}. Parallax distances (with their uncertainties) are given where available from \citet[][and references therein]{Reid2014}, with the kinematic distance in parentheses for comparison. All kinematic distances are the near distance, except for G45.47+0.05 and G49.27$-$0.34 (which are in the direction of tangent points); for the former source we use the parallax distance for G45.45+0.05 which is an \HII\/ region $1\arcmin$ west of the EGO. }    
\tablenotetext{c}{Except for G08.67$-$0.35, this is the Table number of the
  EGO in \citet{egocat}. In that paper, Tables 1 \& 2 list ``likely'' EGOs for which 5-band (3.6 to 24 \mum\/) or only 4.5 \mum\/ {\em Spitzer} photometry can be measured, respectively. Table 4 lists ``possible'' EGO outflow candidates for which only 4.5 \mum\/ photometry is possible. G08.67$-$0.35 is from \citet{Chen2013}.  }
\tablenotetext{d}{Coincidence of EGO with IRDC as indicated by \citet{egocat}, except G08.67-0.35 \citep{Chen2013}.}
\tablenotetext{e}{Water maser data from the \citet{Cyganowski2013} Nobeyama 45-m survey of EGOs, except G08.67-0.35, which comes from \citet{Hofner1996} (VLA) and \citet{Breen2011} (ATCA).}
\tablenotetext{f}{Sources for which we could find no information in the literature are indicted by ``?".}
\tablenotetext{g}{The 6.7~GHz maser detection information comes from \citet{Cyganowski2009} using the VLA, except for G08.67$-$0.35, G12.68$-$0.18, G12.91$-$0.03, G14.63$-$0.58, G16.59$-$0.05, and G45.47$+$0.05 which come from \citet[][and references therein]{Green2010} and used the ATCA.}
\tablenotetext{h}{Information for 44~GHz masers come from the VLA and were taken from \citet[][]{Cyganowski2009}, except for G08.67-0.35 \citep{Gomez2010} and G45.47+0.05 \citep{Kang2015}. }
\tablenotetext{i}{Most information for 95~GHz masers was taken from \citet{Chen2011} using the Mopra 22~m telescope. The exceptions are G08.67$-$0.35 and G14.33$-$0.64 from \citet{Valtts2000} using Mopra, G35.03+0.35 from \citet{Kang2015} using the Korean VLBA Network, and G16.59$-$0.05 and G49.27$-$0.34 from \citet{Chen2012} using the Purple Mountain Observatory 13.7~m telescope. }
\label{distances}
\end{deluxetable*}

Massive young stellar objects (MYSOs) remain embedded in their parent clouds during the early stages of their evolution, making them difficult to observe directly. MYSOs also evolve more quickly than lower-mass young stellar objects (YSOs), making MYSOs rare and especially difficult to observe during their early stages of evolution. Furthermore, MYSOs frequently form in clustered environments, leading to a confusion problem, and tend to be at large distances (>1~kpc), leading to resolution limitations.  One particularly crucial stage of MYSO evolution is the phase in which the object is actively accreting matter and driving outflows. While the process of mass accretion for low-mass stars is fairly well understood \citep{Yorke93}, it is thought that MYSOs continue to accrete even after hydrogen burning has commenced in the core \citep{Stahler00}, and it is this continued mass transfer onto the protostar that leads to the formation of massive stars \citep{Zinnecker07}. However, this process of mass transfer, being as it is both heavily obscured and comparatively short-lived, is not observationally well-constrained.

Recent observations have aimed to investigate the observational markers of MYSOs in this critical phase of their evolution. \citet{egocat} identified >300 sources with extended 4.5 \mum\/ emission in the GLIMPSE-I survey images  \citep{GLIMPSE,Churchwell2009}; these extended 4.5 \mum\/ soures are strongly correlated with infrared dark clouds (IRDCs) and 6.7~GHz Class II \methanol\/ masers. The 4.5~\mum\/ sources were classified as Extended Green Objects (EGOs) by \citet{egocat}, for the common coding of the 4.5~\mum\/ band as green in three-color composite {\it Spitzer} InfraRed Array Camera (IRAC) images. EGOs lie in a region of mid-infrared (MIR) color-color space consistent with protostars that are still in infalling envelopes; the extended ``green'' emission is thought to arise from shocked H$_2$ emission in the 4.5~\mum\/ band.  Furthermore, because IRDCs mark the earliest stages of high-mass star formation \citep{Rathborne07,Rathborne06}, and 6.7~GHz Class II \methanol\/ masers are radiatively pumped and associated exclusively with massive YSOs \citep{Cragg1992,Szymczak_2005,Ellingsen_2006}, \citet{egocat} concluded that EGOs must trace massive protostars that are actively accreting and driving outflows.
From the identified >300 EGO sources, a sample of $\sim$20 objects was selected for follow-up observations in the Class I 44~GHz and Class II 6.7~GHz \methanol\/ maser lines and in the outflow tracers HCO$^+$ and SiO \citep{Cyganowski2009}.  Class I 44 GHz \methanol\/ masers were detected towards 90\% of the sample.  Both the HCO$^+$ line profiles and SiO detections indicated the presence of active outflows in much of the sample, supporting the idea that Class I masers, which are primarily collisionally pumped, trace the impact of outflows on dense gas in star-forming regions \citep[e.g.][]{Plambeck1990,Johnston92,Kurtz2004,Voronkov06}.

Originally discovered in Orion-KL \citep{Barrett71,Barrett75}, the Class~I \methanol\/ $J_2-J_1$-$E$ transitions at 25~GHz form a ladder with energy levels from $\sim$20-140~K for $J$=2-10 \citep[for a rotational level diagram, see][]{Leurini16}. Higher resolution observations soon confirmed the suspicion that the emission in these lines arises from maser action \citep{Hills75}. The first interferometric studies ($J$=6 and 7) noted a correspondence in the maser positions with 2~\mum~H$_2$ emission, thus associating them with shocked gas \citep{Matsakis80}. Further studies of these transitions in other objects find that their intensity usually peaks around $J$=6, and that they are not always inverted but do consistently trace regions of high density and temperature \citep{Menten86,Menten88}.  Statistical equilibrium calculations using the large velocity gradient approach confirm that in gas at $\sim$200~K the $J$=6 maser can occur at densities of $5\times10^{(5-8)}$~cm$^{-3}$  
\citep{Leurini16}.  The 25~GHz transitions are thought to probe a similar, but narrower, range of physical conditions compared to the other two families of Class~I \methanol\/ masers \citep[44/95 and 36/84~GHz,][]{Sobolev2007}.  Thus, interferometric observations of the 25~GHz lines combined with interferometric observations of other Class~I transitions \citep[e.g.][]{Voronkov07,Voronkov12} are important to further refine the physical conditions that the 25~GHz lines typically trace.  

In this paper, we present a 1.3~cm Karl G. Jansky Very Large Array (VLA) survey of 20 GLIMPSE Extended Green Objects (EGOs) in continuum and several \methanol\/ transitions.  The majority of our targets are selected from the GLIMPSE-I EGO catalog of \citet{egocat}; only one, G08.67$-$0.35, is in the GLIMPSE-II survey area \citep[this source is G08.67$-$0.36 in the GLIMPSE-II EGO catalog of][]{Chen2013}.  Table~\ref{distances} summarizes salient information about our target EGOs. 
We describe the observations in Section \S~\ref{observationsVLA}, present our results in Section \S~\ref{results}, discuss the results in Section \S~\ref{discussion}, and summarize our conclusions in Section \S~\ref{conclusions}.

\section{1.3~cm (25~GHz) VLA Observations} \label{observationsVLA}

\begin{deluxetable*}{lcccccccc}
\tablecaption{Observing Parameters} 
\tablefontsize{\scriptsize}
\tablehead{
\colhead{Source} & \multicolumn{2}{c}{Pointing Center (J2000)} & \colhead{Config.}  & \colhead{Date} & \colhead{Phase Cal.} & \colhead{Synth. Beam\tablenotemark{a}} & \colhead{Line rms\tablenotemark{b}} & \colhead{Cont.\ rms}\\
  & \colhead{RA} & \colhead{Dec} & & & & \colhead{$''\times ''$ [P.A.($\arcdeg$)]} & \colhead{(\mjb\/)} & \colhead{(\mjb\/)}
  }
\startdata  
G08.67$-$0.35 & 18 06 18.3 & $-$21 37 31 & CnB & 2011 Jan 22 \& Feb 05 & J1820-2528 & $0.81\times 0.59$ [65.4] & 1.26 & 0.12\\ 
G10.29$-$0.13 & 18 08 49.3 & $-$20 05 57 & D   & 2010 Sep 13 & J1820-2528 & $4.52\times 2.38$ [14.0] & 5.82 & 0.31\\
G10.34$-$0.14 & 18 09 00.0 & $-$20 03 35 & D   & 2010 Sep 09 & J1820-2528 & $5.15\times 2.51$ [21.4] & 6.50 & 0.29\\
G11.92$-$0.61 & 18 13 58.1 & $-$18 54 17 & D \& CnB & 2010 Aug 25 \& 2011 Jan 30 & J1820-2528 & $1.30\times 0.87$ [-7.3] & 2.88 & 0.08\\
G12.68$-$0.18 & 18 13 54.7 & $-$18 01 47 & CnB & 2011 Jan 29 & J1832-2039 & $0.87\times 0.54$ [71.0] & 1.54 & 0.06\\
G12.91$-$0.03 & 18 13 48.2 & $-$17 45 39 & C   & 2010 Dec 11 & J1832-2039 & $1.47\times 0.86$ [-176.0] & 2.80 & 0.05\\
G14.33$-$0.64 & 18 18 54.4 & $-$16 47 46 & D   & 2010 Sep 10 & J1832-2039 & $4.63\times 2.45$ [20.0] & 5.20 & 0.10\\
G14.63$-$0.58 & 18 19 15.4 & $-$16 30 07 & D   & 2010 Sep 12 & J1832-2039 & $4.42\times 2.50$ [11.4] & 3.73 & 0.06\\
G16.59$-$0.05 & 18 21 09.1 & $-$14 31 48 & C   & 2011 Jan 17 & J1832-1035 & $1.37\times 0.89$ [-0.9] & 3.54 & 0.07\\
G18.67$+$0.03 & 18 24 53.7 & $-$12 39 20 & C   & 2011 Jan 07 & J1832-1035 & $1.31\times 0.81$ [-2.4] & 2.03 & 0.04\\
G18.89$-$0.47 & 18 27 07.9 & $-$12 41 36 & C   & 2010 Dec 31 & J1832-1035 & $1.36\times 0.82$ [-4.0] & 2.89 & 0.06\\
G19.36$-$0.03 & 18 26 25.8 & $-$12 03 57 & D   & 2010 Aug 22 & J1832-1035 & $4.62\times 2.67$ [22.6] & 6.43 & 0.11\\
G22.04$+$0.22 & 18 30 34.7 & $-$09 34 47 & D   & 2010 Aug 30 & J1832-1035 & $4.02\times 2.64$ [19.0] & 6.60 & 0.11\\
G24.94$+$0.07 & 18 36 31.5 & $-$07 04 16 & D   & 2010 Sep 03 & J1832-1035 & $4.05\times 2.78$ [26.7] & 3.90 & 0.09\\
G25.27$-$0.43 & 18 38 56.9 & $-$07 00 48 & C   & 2011 Jan 06 & J1832-1035 & $1.27\times 1.05$ [19.5] & 3.73 & 0.05\\
G28.28$-$0.36 & 18 44 13.2 & $-$04 18 04 & D   & 2010 Sep 05 & J1832-1035 & $3.18\times 2.33$ [-4.8] & 5.31 & 0.17\\
G28.83$-$0.25 & 18 44 51.3 & $-$03 45 48 & C   & 2011 Jan 08 & J1851+0035 & $1.16\times 0.83$ [-8.8] & 2.29 & 0.05\\
G35.03$+$0.35 & 18 54 00.5 & $+$02 01 18 & D   & 2010 Sep 07 & J1851+0035 & $3.75\times 2.81$ [-51.9] & 4.51 & 0.10\\
G45.47$+$0.05 & 19 14 25.6 & $+$11 09 28 & C   & 2010 Dec 12 \& 24 & J1922+1530 & $0.99\times 0.82$ [-19.7] & 1.62 & 0.09\\
G49.27$-$0.34 & 19 23 06.7 & $+$14 20 13 & C   & 2010 Dec 19 & J1922+1530 & $0.95\times 0.83$ [-34.4] & 2.90 & 0.07
\enddata
\tablenotetext{a}{Synthesized beam of the CH$_3$OH-E 5(2,3)-5(1,4) transition.}
\tablenotetext{b}{Median rms noise per channel in the CH$_3$OH-E 5(2,3)-5(1,4) image cubes. The rms noise in a channel with bright maser emission will be significantly higher due to dynamic range limitations.}
\label{observationsTable}
\end{deluxetable*}

We used the VLA \citep{Perley2011} to observe 20 EGOs at 1.3~cm (25~GHz). Table \ref{observationsTable} summarizes the project AB1346 phasecenters, observing dates, configuration(s), and phase calibrator for each target EGO.  The observations were taken under the Resident Shared Risk Observing (RSRO) program \citep{Chandler2014} using $16\times 8$~MHz spectral windows (each with 256 channels and single polarization) to observe four transitions of CH$_3$OH, as well as the \ammonia\/ (1,1) through (6,6) metastable transitions and the H63$\alpha$ and H64$\alpha$ radio recombination lines (RRLs). The four remaining spectral windows were placed to cover additional possible, but unlikely to be detected, species of interest. The primary purpose of these ``extra'' spectral windows is for continuum, and indeed, none of these transitions were detected. In this paper we focus on the 1.3~cm continuum and  \methanol\/ data; the details of the observed \methanol\/ transitions are given in Table~\ref{lines}. Hereafter, the \methanol\/ transitions will be denoted by the first two values of their upper state quantum number, for example J$^\prime$(K$_a$,K$_c$) - J$^{\prime \prime}$(K$_a$,K$_c$) = 3(2,1) - 3(1,2) will be 3$_2$, etc. 

The data were calibrated and imaged using the CASA software package. For all sources, the bandpass calibrator was J1924$-$2914.  For all but two sources, 3C286 (J1331+3030), combined with a model for its flux distribution, was used for absolute flux calibration. The two exceptions (where the 3C286 observations failed to provide viable data) were G16.59$-$0.05 and G35.03$+$0.35.  For these two sources, the derived flux density for the nearest other observation of the same phase calibrator (in time) was used to set the absolute flux scale.  Opacities as a function of frequency were derived from the VLA seasonal model\footnote{See EVLA Memo 143, VLA Test Memo 232, and VLA Scientific Memo 176. All three memos are archived at http://library.nrao.edu/vla.shtml.}. 
We expect the absolute flux calibration to be good to $\sim 10\%$. Where necessary, antenna position corrections were also applied.

After the standard calibration was applied, ``line'' datasets were created by removing the continuum in the uv-plane using line-free channels in each spectral window.  A few of the EGOs have bright, compact continuum sources in the VLA field of view (FOV) that are not at the phase center: G08.67$-$0.35, G11.92$-$0.61, and G28.28$-$0.36. In these cases, it was necessary to first shift the phase center to the brightest continuum source in the field of view, and then shift back after continuum subtraction to avoid aliasing effects.  After continuum subtraction, the 31.25 kHz channel width ($\sim 0.38$ \kms\/) line data were Hanning smoothed and imaged with a velocity channel width of 0.4~\kms\/. The D-configuration sources were imaged with a robust parameter of 0.75, while the CnB and C configuration data (see Table~\ref{observationsTable}) were imaged with robust=1.0. 

The continuum for each EGO comprises 30 MHz of bandwidth from the four ``extra'' transition spectral windows, plus an additional 15 MHz from the line-free regions of the spectral windows covering the four \methanol\/ transitions. Fields without RRL detections in the FOV have an additional 7.5~MHz of continuum bandwidth. The continuum images were made with multi-frequency synthesis and robust=1.0 for targets with only weak continuum emission in the FOV, and more uniform weighting and/or a restricted short-spacing uv-range when diffuse/confusing sources are present. The median geometric means of the synthesized beam in the D, C, and CnB  configurations are 3$\farcs$32, 1$\farcs$03, and 0$\farcs$69, respectively (the source with D \& CnB configuration data is included in the CnB median). 
The imaged fields of view for both the line and continuum images are similar to the $2'$ full width to half power (FWHP) of the 25m VLA dishes at 1.3~cm; primary beam correction was applied to all images. 



\begin{deluxetable}{lcccc}
\tablecaption{Observed CH$_3$OH Transitions\tablenotemark{a}} 
\tabletypesize{\footnotesize}
\tablehead{
\colhead{Species} & \colhead{Resolved QNs} & \colhead{Frequency\tablenotemark{b}} & \colhead{$E_{upper}$} & \colhead{S$_{ij}$ $\mu^{2}$} \\
 & & \colhead{(GHz)} & \colhead{(K)} & \colhead{(D$^2$)} 
}
\startdata
CH$_3$OH-$E$ & 3(2,1)-3(1,2) & 24.928707(7) & 36.17 & 2.8073\\
CH$_3$OH-$E$ & 5(2,3)-5(1,4) & 24.9590789(4) & 57.07 & 5.0264\\
CH$_3$OH-$E$ & 8(2,6)-8(1,7) & 25.2944165(2)  & 105.84 & 8.3910\\
CH$_3$OH-$E$ & 10(2,8)-10(1,9) & 25.8782661(4) & 149.97 & 10.7398
\enddata
\tablenotetext{a}{Transition properties taken from \citet{Muller04}.}
\tablenotetext{b}{Numbers in parentheses denote the measurement uncertainties in units of the least significant figure.}
\label{lines}
\end{deluxetable}

\section{Results}
\label{results}

\begin{figure*}
    \figurenum{1}
    \centering
    \includegraphics[width=0.9\textwidth]{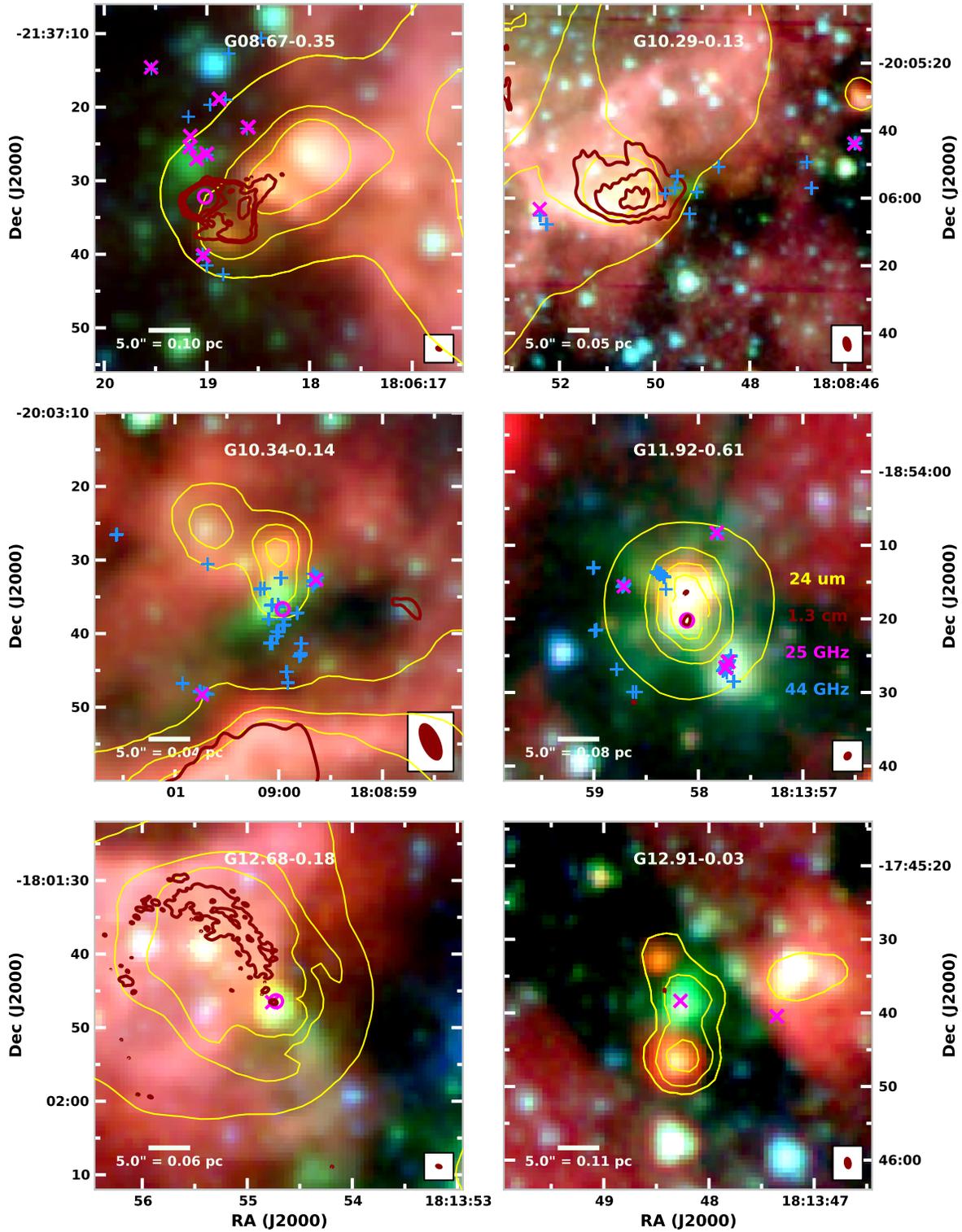}
    \caption{{\it Spitzer} GLIMPSE 3-color images (RGB: 8.0, 4.5, and 3.6 $\mu$m) for sources with detected 25~GHz \methanol\/ emission.  For each EGO, the displayed FOV is centered on the coordinates given in Table \ref{observationsTable}. {\it Spitzer} MIPSGAL 24~$\mu$m contours are overlaid in yellow (contour levels, in MJy sr$^{-1}$: G08.67: (300, 800, 1600); G10.29: (1200, 1800); G10.34: (900, 1300, 1700); G11.92: (600, 1200, 1800); G12.68: (900, 1300); G12.91: (300, 600, 1200); G14.33 (1000, 1500, 2000); G14.63: (300, 600); G16.59: (300, 600, 1200); G18.67: (300, 900); G19.36: (300, 600, 1400); G22.04: (300, 900); G24.94: (300, 600, 900); G28.28: (800, 1600); G35.03: (900, 1300, 1700); G45.47: (800, 1600)).  VLA 1.3~cm continuum contours are overlaid in dark red (levels: 4, 12, 28, 60$\times\sigma$, where $\sigma$ for each field is given in Table \ref{observationsTable}).  Sites of 25~GHz \methanol\/ maser emission are marked by magenta $\times$ symbols, while sites of thermal 25~GHz \methanol\/ emission are marked by magenta $\circ$ symbols.  Class I 44~GHz \methanol\/ masers from the literature (where available) are marked with blue $+$ symbols (see Table \ref{distances}).}
    \label{3color}
\end{figure*}

\begin{figure*}
    \figurenum{1}
    \centering
    \includegraphics[width=0.9\textwidth]{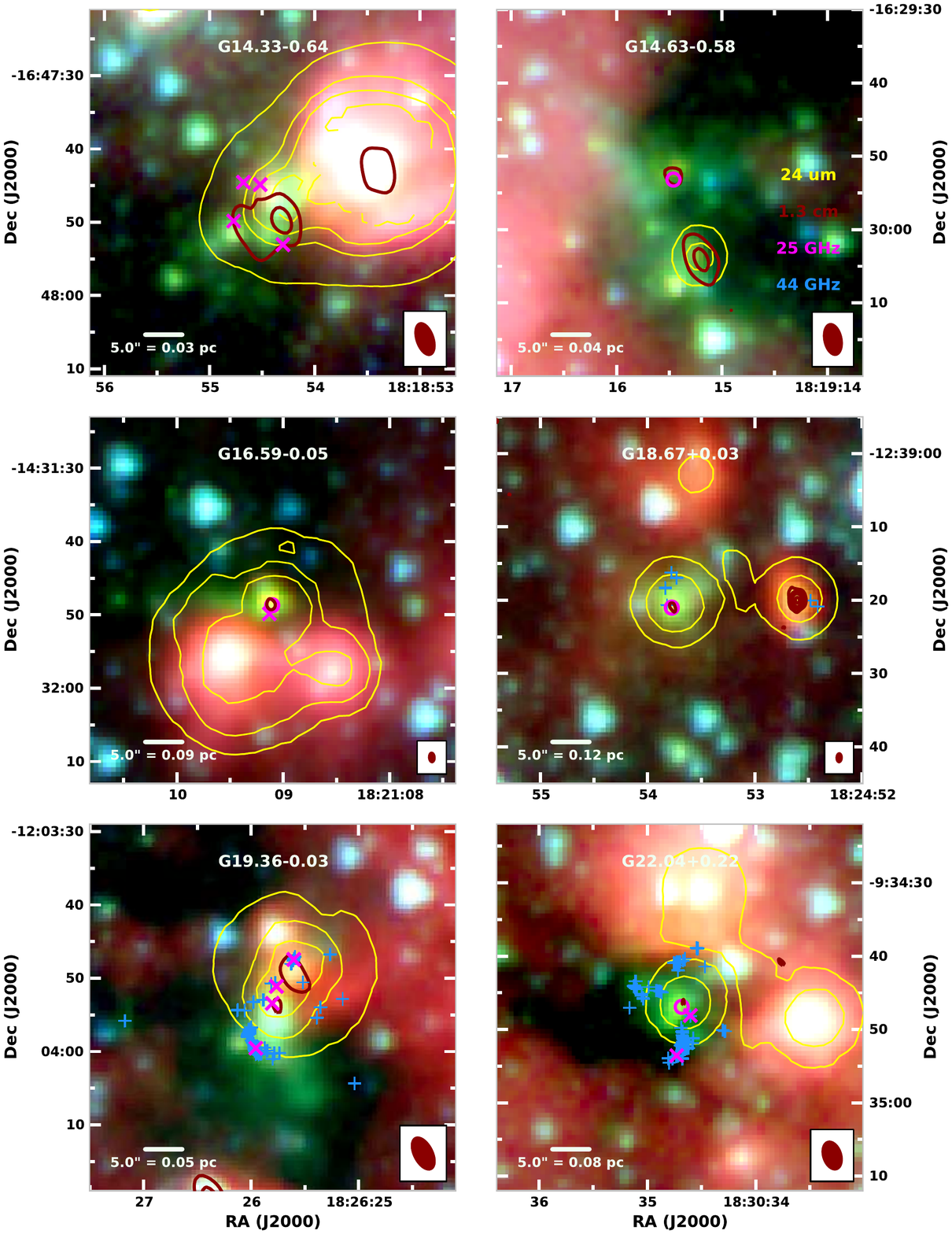}
    \caption{cont'd.}
\end{figure*}

\begin{figure*}
    \figurenum{1}
    \centering
    \includegraphics[width=0.9\textwidth]{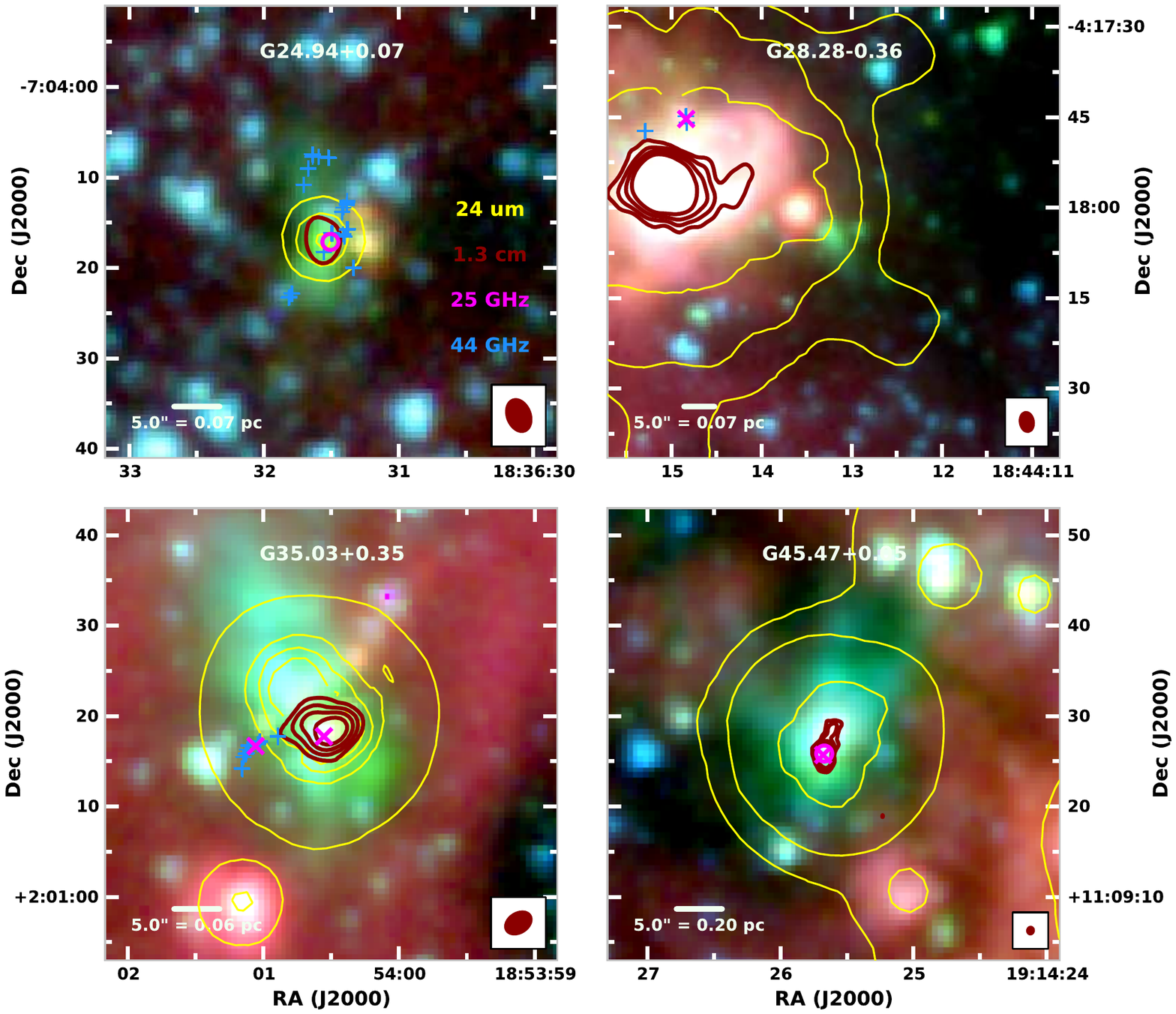}
    \caption{cont'd.}
\end{figure*}

\begin{figure*}
    \figurenum{2}
    \centering
    \includegraphics[width=0.9\textwidth]{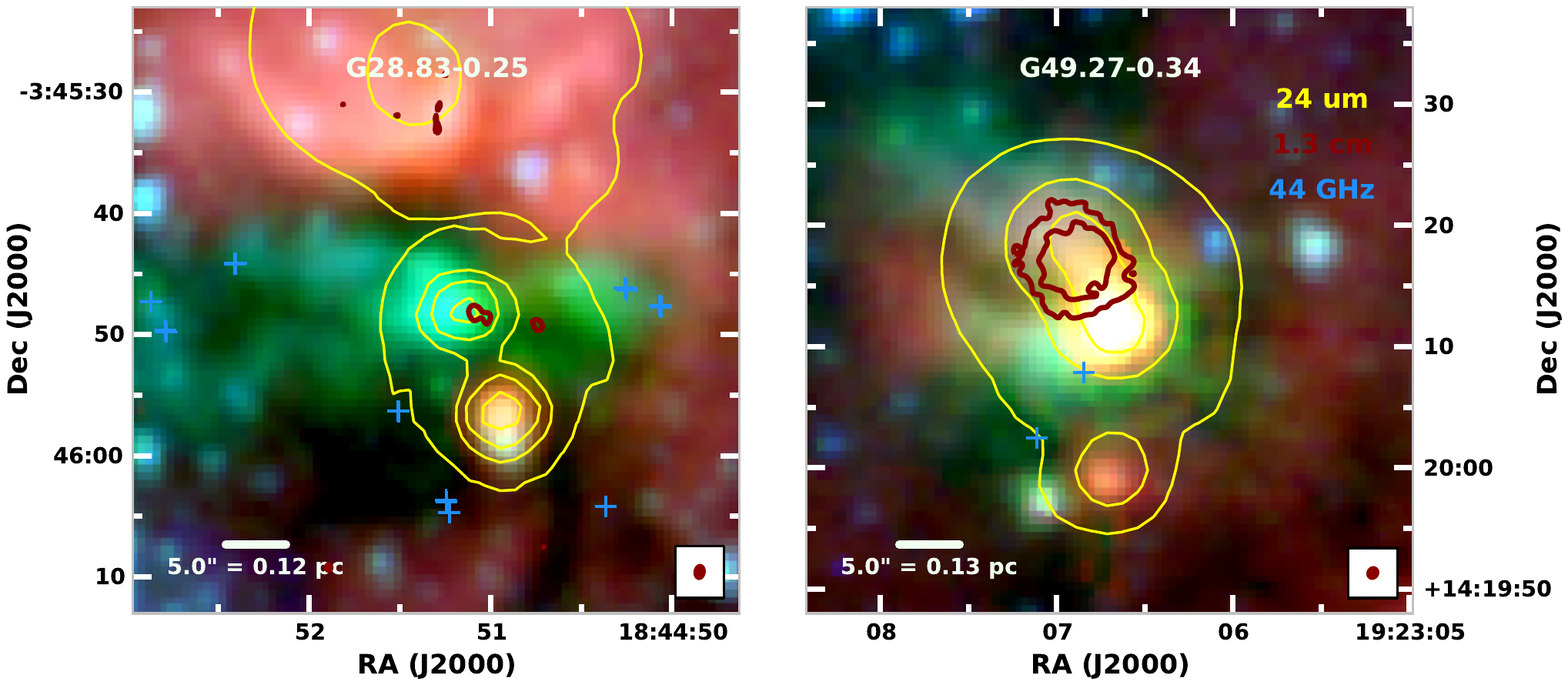}
    \caption{Same as Fig.~\ref{3color}, but for sources with 1.3~cm continuum detections but no detected 25~GHz \methanol\/ emission. MIPSGAL 24 \mum\/ contour levels are: G28.83: 900, 1300, 1700 MJy sr$^{-1}$; G49.27: 800, 1600 MJy sr$^{-1}$.}
    \label{cmonly}
\end{figure*}


Figures~\ref{3color} and \ref{cmonly} show three-color mid-infrared {\em Spitzer} GLIMPSE \citep{GLIMPSE,Churchwell2009} images of our EGO targets. In both figures, contours of the new VLA 1.3~cm continuum data are overlaid in red. We also overlay 24~\mum\/ contours from the MIPS/{\em Spitzer} Survey of the Galactic Plane (MIPSGAL) \citep{MIPSGAL} in yellow. Fig.~\ref{3color} shows the 16 targets for which we detected 25~GHz \methanol\/ emission. 
Fig.~\ref{cmonly} shows the two targets for which we detected 1.3~cm continuum emission in the vicinity of the EGO but no 25~GHz \methanol\/ emission. Note that the two sources (G18.89$-$0.47 and G25.27$-$0.43) for which neither 1.3~cm continuum nor 25~GHz \methanol\/ emission was detected in the vicinity of the EGO are not shown.

\subsection{1.3~cm Continuum Emission}
\label{cmresults}

\begin{deluxetable*}{lcccccccc}[ht]
\tablecaption{1.3 cm Continuum Flux Densities For Emission Associated with EGO Targets}
\tabletypesize{\footnotesize}
\tablehead{\colhead{Source} & \multicolumn{2}{c}{J2000 Coordinates} & \colhead{Peak intensity} & \colhead{Integrated$^a$} & \colhead{Prior 1.3~cm} & \colhead{Ref.$^b$} & \colhead{Prior cm-$\lambda$} & \colhead{Ref.}\\
\colhead{Name} & \colhead{$\alpha^{(h}$ $^m$ $^{s)}$} & \colhead{$\delta$($^{\circ}$ $\arcmin$ $\arcsec$)} & \colhead{(mJy beam$^{-1}$)} & \colhead{(mJy)} & \colhead{Detections?} & & \colhead{Detections?} &
}
\startdata
G08.67$-$0.35\_CM1 & 18:06:19.01 & -21:37:32.2 & 183.0 (0.1) & 1141 (1) & N & -- & Y & W89\\
G10.29$-$0.14\_CM1 & 18:08:50.47 & -20:06:01.5 & 10.7 (0.3) & 201 (2) & Y & C11b$^c$ & Y & C11b$^c$\\
G11.92$-$0.64\_CM1 & 18:13:58.108 (0.003) & -18:54:20.212 (0.09) & 0.62 (0.08) & unr & Y & C11b & Y & M16\\
~~~~~~~~~~~~~~~~~~~~~\_CM2 & 18:13:58.121 (0.009) & -18:54:16.46 (0.13) & 0.32 (0.07) & unr & N & -- & Y & C17\\
G12.68$-$0.18\_CM1 & 18:13:54.750 (0.002) & -18:01:46.52 (0.01) & 0.73 (0.06) & unr & Y & M16 & Y & M16\\
G12.91$-$0.03\_CM1 & 18:13:48.41 (0.01) & -17:45:36.92 (0.2) & 0.23 (0.05) & unr & N & -- & N & --\\
G14.33$-$0.64\_CM1 & 18:18:54.30  & -16:47:50.0 & 1.78 (0.1) & 4.1 (0.3) & N & -- & N & --\\
G14.63$-$0.58\_CM1 & 18:19:15.19 & -16:30:04.0 & 0.99 (0.06) & 1.2 (0.1) & N & -- & N & --\\
~~~~~~~~~~~~~~~~~~~~~\_CM2 & 18:19:15.469 (0.03) & -16:29:52.4 (0.4) & 0.29 (0.07) & unr & N & -- & N & --\\
G16.59$-$0.05\_CM1 & 18:21:09.116 (0.004) & -14:31:48.58 (0.08) & 0.56 (0.07) & unr & Y & R16, H11$^d$, Z06 & Y & M16, R16, Z06 \\
G18.67$+$0.03\_CM1 & 18:24:53.755 (0.007) & -12:39:20.9 (0.2) & 0.18 (0.04) & unr & N & -- & N & --\\
G19.36$-$0.03\_CM1 & 18:26:25.60 (0.02) & -12:03:49.6 (0.4) & 0.7 (0.1) & 1.3 (0.3) & N & -- & Y & C11b$^e$\\
~~~~~~~~~~~~~~~~~~~~~\_CM2 & 18:26:25.77 (0.01) & -12:03:53.7 (0.5) & 0.5 (0.1) & unr & N & -- & N & --\\
G22.04$+$0.22\_CM1 & 18:30:34.70 (0.03) & -09:34:46.2 (0.5) & 0.6 (0.1) & unr & N & -- & N & --\\
G24.94$+$0.07\_CM1 & 18:36:31.563 (0.008) & -07:04:16.8 (0.2) & 0.85 (0.09) & unr & N & -- & Y & C11b\\
G28.83$-$0.25\_CM1 & 18:44:50.74 (0.01) & -03:45:49.19 (0.08) & 0.24 (0.04) & unr & N & -- & Y & C11b\\
~~~~~~~~~~~~~~~~~~~~~\_CM2 & 18:44:51.086 (0.005) & -03:45:48.22 (0.08) & 0.35 (0.05) & unr & N & -- & Y & C11b\\
G35.03$-$0.03\_CM1 & 18:54:00.50 & +02:01:18.5 & 13.9 (0.1) & 18.5 (0.3) & Y & B11, C11b$^f$ & Y & C11b\\
G45.47$+$0.05\_CM1 & 19:14:25.68 & +11:09:25.8 & 110.09 (0.09) & 180.9 (0.4) & Y & H99 & Y & W89\\
G49.27$-$0.34\_CM1 & 19:23:06.87 & +14:20:18.2 & 1.62 (0.07) & 57.8 (0.6) & Y & C11b & Y & C11b
\enddata
\tablenotetext{a}{Flux densities with reported values were measured within the $3\sigma$ contour; see Table~\ref{observationsTable} for $\sigma$ values. In this case, the position is that of the peak pixel within the $3\sigma$ contour. Flux densities designated by `unr' indicate that the emission is unresolved. For these cases, the Gaussian fitted position and peak intensity along with the statistical uncertainties are reported.}
\tablenotetext{b}{Reference abbreviations correspond to: B11: \citet{Brogan11}. C11b: \citet{Cyganowski2011b}. C17: \citet{Cyganowski2017}. H99: \citet{Hofner1999}. H11: \citet{Hofner2011}. M16: \citet{Moscadelli2016}. R16: \citet{Rosero16}. W89: \citet{Wood1989}. Z06: \citet{Zapata2006}.}
\tablenotetext{c}{\citet{Cyganowski2011b} list this as a field source, designation F-CM1.}
\tablenotetext{d}{\citet{Hofner2011} list this source as IRAS 18182-1433, Component B.}
\tablenotetext{e}{\citet{Cyganowski2011b} list this is a field source, designation F-CM1.}
\tablenotetext{f}{G35.03$-$0.03\_CM1 from the current work is a blend of CM1, CM2, CM4, and CM5 from \citet{Cyganowski2011b}.}
\label{cm_fluxes}
\end{deluxetable*}

We detect 1.3~cm continuum emission within the $2'$ field of view for 19 of the 20 observed fields (only G18.89$-$0.47 lacks any detectable emission). However, detectable 1.3~cm emission in the vicinities of the EGOs themselves (i.e. within or touching the boundary of the extended 4.5~\mum\/ emission) is only detected toward 16 of the 20 fields for a detection rate of $80\%$ (Figs.~\ref{3color} and \ref{cmonly}). A few of these have more than one distinct region of emission, bringing the total number of individual EGO-associated 1.3~cm detections to 20. The positions and properties of the EGO-associated 1.3~cm continuum detections are detailed in Table~\ref{cm_fluxes}. 

Despite the modest aggregate continuum bandwidth obtainable from the relatively narrow spectral windows (see \S~\ref{observationsVLA}), in many cases these data represent the most sensitive cm wavelength (cm-$\lambda$) observations of these sources to date. For example, twelve EGOs in our sample were also included in the 1.3 and 3.6~cm VLA EGO continuum survey of \citet[][resolution $\sim$1\arcsec]{Cyganowski2011b}. With the exception of G10.29$-$0.13 (which is severely dynamic-range-limited by emission from a bright \HII\/ region in the field), our new  1.3~cm images are a factor of 2-4 more sensitive than those of \citet{Cyganowski2011b}.  

Of our 20 EGO-associated continuum detections, 13 are new detections at 1.3~cm and have a median peak intensity of 0.50~\mjyb. Of these, 7 are new detections at any cm-$\lambda$. Columns 6-9 of Table \ref{cm_fluxes} list previous detections of each source at 1.3~cm and other cm-$\lambda$. References for previously-detected sources, as well as alternate names where applicable, are listed in columns 7 and 9 and associated table notes. It is notable that if it had been previously observed at 1.3~cm, it is very likely that G08.67$-$0.35\_CM1 would have been detected at past sensitivity levels.
The other 12 new 1.3~cm detections, with a median peak intensity of 0.42 mJy beam$^{-1}$, are sufficiently weak that these are the first (published) data with the sensitivity to detect them. Sub-mJy emission at 1.3~cm (at kiloparsec distances) can be due to either free-free emission from protostellar winds/jets or hypercompact \HII\/ regions, or alternatively the Rayleigh-Jeans tail of dust emission \citep[see for example][]{Brogan2016}. In \S\ref{individual}, we discuss the morphology of the 1.3~cm continuum emission and its relationship to the target EGO for individual sources, and compare with other cm-$\lambda$ data where possible. Unfortunately, the absence (for the most part) of data at a second cm-$\lambda$ with resolution and sensitivity comparable to our 1.3~cm images precludes systematic analysis of the underlying emission mechanism(s).

\subsection{25~GHz \methanol\/ Emission}
\label{methanol}

\begin{turnpage}
\begin{deluxetable*}{lccccccccccc}
\tablecaption{25 GHz CH$_3$OH Emission: Fitted and Derived Properties}
\tablecolumns{11}
\tabletypesize{\footnotesize}
\tablehead{
 & \multicolumn{2}{c}{Fitted Position (J2000)} & & & \colhead{Fitted Size\tablenotemark{a}} &  \multicolumn{4}{c}{Fitted Flux Densities\tablenotemark{b}} &  & \\
\colhead{Emission} & \colhead{RA} & \colhead{Dec} & \colhead{V$_{peak}$\tablenotemark{c}} & \colhead{V$_{range}$\tablenotemark{d}} & \colhead{Major $\times$ Minor [PA]} & \colhead{3$_2$ Flux} & \colhead{5$_2$ Flux} & \colhead{8$_2$ Flux} & \colhead{10$_2$ Flux} & \colhead{5$_2$ $T_B$} & \colhead{Emission}\\
 \colhead{Region} &  \colhead{($^{\rm h}$ $^{\rm m}$ $^{\rm s}$)} & \colhead{($^{\circ}$ $\arcmin$ $\arcsec$)} & \colhead{(\kms\/)} & \colhead{(\kms\/)} & \colhead{($''$) $\times$ ($''$) [$^{\circ}$]} & \colhead{(mJy)} & \colhead{(mJy)} & \colhead{(mJy)} & \colhead{(mJy)} & \colhead{(K)} & \colhead{Type\tablenotemark{e}}
}
\startdata
G08.67-0.35\_a & 18:06:18.596 (0.005) & -21:37:22.72 (0.09) & 38.2 & 37.0 - 39.0 & 0.72$\times$<0.69 (0.27) & -- & 10 (3) & -- & -- & 40 & Maser\\
G08.67-0.35\_b & 18:06:18.881 (0.002) & -21:37:18.90 (0.02) & 38.2 & 37.4 - 39.0 & <0.69 & --  & 13 (2) & 5 (1) & -- & 52 & Maser\\
G08.67-0.35\_c & 18:06:19.003 (0.001) & -21:37:26.34 (0.01) & 35.0 & 34.2 - 35.8 & 0.39$\times$<0.69 (0.09) & 18 (3) & 44 (2) & 6 (1) & -- & 320 & Maser\\
G08.67-0.35\_d & 18:06:19.019 (0.004) & -21:37:32.15 (0.05) & 36.2 & 35.4 - 39.0 & 1.63$\times$1.41 [80] (0.18$\times$0.16 [77]) & 120 (10) & 130 (10) & 63 (6) & 40 (7) & 110 & Thermal+Abs\\
G08.67-0.35\_e & 18:06:19.036 (0.001) & -21:37:40.18 (0.01) & 33.4 & 32.6 - 34.6 & 0.27$\times$<0.34* (0.06) & 23 (5) & 99 (3) & 31 (2) & -- & 2110 & Maser\\
G08.67-0.35\_f & 18:06:19.101 (0.002) & -21:37:27.05 (0.02) & 36.2 & 35.8 - 37.0 & 0.40$\times$<0.69 (0.15) & -- & 23 (3) & 15 (2) & -- & 165 & Maser\\
G08.67-0.35\_g & 18:06:19.162 (0.001) & -21:37:23.95 (0.10) & 35.4 & 35.0 - 36.6 & <0.69 & -- & 28 (2) & -- & -- & 116 & Maser\\
G08.67-0.35\_h & 18:06:19.168 (0.0001) & -21:37:25.427 (0.002) & 35.8 & 34.2 - 37.4 & 0.15$\times$0.10 (0.05$\times$0.04)  [35 (45)] & 106 (4) & 791 (9) & 91 (3) & 14 (4) & 104000 & Maser\\
G08.67-0.35\_i & 18:06:19.544 (0.001) & -21:37:14.60 (0.01) & 36.6 & 35.8 - 37.4 & <0.69 & -- & 27 (2) & -- & -- & 113 & Maser\\
\hline

G10.29-0.13\_a & 18:08:45.802 (0.001) & -20:05:43.85 (0.04) & 14.6 & 13.8 - 15.0 & 1.31$\times$<1.64* (0.37) & 41 (9) & 288 (7) & 97 (5) & -- & 263 & Maser\\  
G10.29-0.13\_b & 18:08:52.416 (0.003) & -20:06:03.20 (0.10) & 15.4 & 15.0 - 16.6 & <3.28 & -- & 97 (7) & 60 (10) & -- & 18 & Maser\\
\hline

G10.34-0.14\_a & 18:08:59.640 (0.0002) & -20:03:32.661 (0.01) & 14.8 & 14.0 - 15.6 & 1.32$\times$0.50 [34.8] (0.08$\times$0.05 [3.3]) & 1058 (9) & 2760 (10) & 766 (6) & -- & 8200 & Maser\\
G10.34-0.14\_b & 18:08:59.963 (0.022) & -20:03:36.68 (0.39) & 12.8 & 9.2 - 13.6  & 5.29$\times$<2.51 (1.78) & 30 (6) & 41 (7) & 37 (7) & -- & 6 & Thermal\\
G10.34-0.14\_c & 18:09:00.738 (0.014) & -20:03:48.30 (0.42) & 11.6 & 11.2 - 12.4 & <3.60 & 14 (3) & 36 (5) & -- & -- & 5.4 & Maser\\
\hline

G11.92-0.61\_a & 18:13:57.715 (0.0001) & -18:54:25.836 (0.01) & 34.0 & 32.8 - 34.4 & 0.50$\times$0.16 [154] (0.04$\times$0.07 [5]) & 150 (3) & 511 (6) & 19 (5) & -- & 12500 & Maser\\
G11.92-0.61\_b & 18:13:57.739 (0.001) & -18:54:26.64 (0.02) & 35.2 & 34.8 - 35.6 & 0.55$\times$<1.06 (0.14) & 18 (3) & 35 (2) & -- & -- & 116 & Maser\\
G11.92-0.61\_c & 18:13:57.824 (0.002) & -18:54:08.30 (0.11) & 36.4 & 35.6 - 37.2 & 1.53$\times$<1.06 (0.38) & -- & 18 (3) & -- & -- & 22 & Maser\\
G11.92-0.61\_d & 18:13:58.109 (0.007) & -18:54:20.21 (0.19) & 36.8 & 33.6 - 40.0 & 3.71$\times$2.01 [9] (0.50$\times$0.27 [9]) & 80 (10) & 54 (7) & 41 (6) & 42 (7) & 14 & Thermal\\
G11.92-0.61\_e & 18:13:58.715 (0.002) & -18:54:15.57 (0.06) & 36.8 & 36.4 - 37.2 & 1.13$\times$0.59 [153] (0.25$\times$0.22 [22]) & 22 (2) & 18 (2) & -- & -- & 53 & Maser\\
\hline

G12.68-0.18\_a & 18:13:54.726 (0.003) & -18:01:46.41 (0.02) & 56.0 & 53.2 - 59.2 & 1.06$\times$0.69 [90] (0.14$\times$0.10 [17]) & 24 (4) & 33 (3) & 39 (3) & 43 (4) & 90 & Thermal\\
G12.68-0.18\_b & 18:13:54.763 (0.002) & -18:01:46.54 (0.05) & 52.4 & 52.0 - 52.8 & 0.91$\times$0.28 [174] (0.18$\times$0.29 [18]) & -- & 18 (2) & 13 (2) & 19 (3) & 140 & Maser$^{*}$\\
\hline

G12.91-0.03\_a & 18:13:47.355 (0.001) & -17:45:40.46 (0.05) & 55.0 & 54.2 - 56.2 & <1.12 & -- & 23 (3) & -- & -- & 36 & Maser\\
G12.91-0.03\_b & 18:13:48.271 (0.001) & -17:45:38.36 (0.04) & 57.8 & 56.6 - 60.2 & 1.25$\times$<1.12 (0.15) & 50 (20) & 59 (3) & 27 (3) & -- & 82 & Maser\\
\hline

G14.33-0.64\_a & 18:18:54.302 (0.006) & -16:47:53.06 (0.16) & 21.6 & 20.8 - 21.6 & 3.24$\times$2.52 [7] (0.73$\times$0.56 [88]) & -- & 99 (8) & -- & -- & 24 & Maser\\
G14.33-0.64\_b & 18:18:54.519 (0.002) & -16:47:44.86 (0.04) & 20.4 & 20.0 - 21.2 & 1.56$\times$1.13 [47] (0.36$\times$0.40 [45]) & 50 (10) & 400 (10) & 362 (7) & 96 (5) & 447 & Maser\\
G14.33-0.64\_c & 18:18:54.678 (0.007) & -16:47:44.54 (0.14) & 22.4 & 22.0 - 23.2 & <2.87 & 8 (2) & 41 (3) & 31 (2) & -- & 7.1 & Maser\\
G14.33-0.64\_d & 18:18:54.771 (0.007) & -16:47:49.84 (0.11) & 22.8 & 22.4 - 23.2 & 3.18$\times$<2.45 (0.32) & 31 (6) & 58 (3) & 52 (5) & 13 (2) & 14.6 & Maser\\
\hline

G14.63-0.58\_a & 18:19:15.46 (0.01) & -16:29:53.18 (0.32) & 19.4 & 19.4 - 19.8 & 5.07$\times$<2.50 (1.11) [14] & 27 (5) & 19 (3) & -- & -- & 2.9 & Thermal\\
\hline

G16.59-0.05\_a & 18:21:09.104 (0.002) & -14:31:48.69 (0.07) & 58.6 & 57.0 - 59.8 & 1.68$\times$0.96 [167] (0.24$\times$0.14 [12]) & 25 (4) & 36 (3) & 27 (3) & 19 (3) & 44 & Thermal\\
G16.59-0.05\_b & 18:21:09.131 (0.0003) & -14:31:49.860 (0.01) & 61.4 & 61.0 - 61.8 & 0.40$\times$0.22 [178] (0.08$\times$0.05 [19]) & 16 (3) & 320 (6) & 205 (4) & 29 (4) & 7200 & Maser\\
\hline

G18.67+0.03\_a & 18:24:53.775 (0.005) & -12:39:21.01 (0.15) & 78.2 & 78.2 - 79.4 & 1.68$\times$0.91 [165] (0.51$\times$0.34 [37]) & -- & 12 (2) & 6 (1) & -- & 15 & Thermal\\
\hline

G19.36-0.03\_a & 18:26:25.604 (0.001) & -12:03:47.44 (0.03) & 26.0 & 25.6 - 26.4 & <1.75* & 60 (10) & 130 (3) & -- & -- & 84 & Maser\\
G19.36-0.03\_b & 18:26:25.765 (0.002) & -12:03:51.113 (0.05) & 26.4 & 25.2 - 26.8 & <3.51 & 29 (3) & 91 (4) & 117 (2) & 38 (1) & 14.4 & Maser$^{*}$\\
G19.36-0.03\_c & 18:26:25.803 (0.003) & -12:03:53.43 (0.16) & 27.6 & 27.6 - 28.0 & <3.51 & -- & 46 (2) & -- & -- & 7.4 & Maser\\
G19.36-0.03\_d & 18:26:25.957 (0.0002) & -12:03:59.547 (0.01) & 26.4 & 24.4 - 28.4 & 1.31$\times$0.52 [30] (0.05$\times$0.03 [2]) & 2870 (10) & 6390 (20) & 1370 (10) & 135 (6) & 18420 & Maser\\
\hline

G22.04+0.22\_a & 18:30:34.603 (0.0002) & -09:34:48.087 (0.01) & 51.8 & 51.0 - 55.0 & <1.63* & 1090 (10) & 7650 (30) & 3920 (10) & 723 (6) & 5660 & Maser\\
G22.04+0.22\_b & 18:30:34.682 (0.005) & -09:34:46.91 (0.13) & 50.2 & 47.8 - 50.6 & 2.71$\times$1.70 [23] (0.58$\times$0.39 [34]) & 70 (10) & 200 (20) & 160 (5) & 70 (4) & 86 & Thermal\\
G22.04+0.22\_c & 18:30:34.729 (0.001) & -09:34:53.56 (0.02) & 50.2 & 49.4 - 50.6 & 0.69$\times$<1.63* (0.30) & 48 (4) & 342 (7) & 47 (7) & -- & 598 & Maser\\
\hline

G24.94+0.07\_a & 18:36:31.506 (0.010) & -07:04:17.12 (0.43) & 42.2 & 39.0 - 45.0 & <3.36 & -- & 11 (3) & -- & -- & 2.0 & Thermal\\
\hline

G28.28-0.36\_a & 18:44:14.836 (0.003) & -04:17:45.29 (0.09) & 48.2 & 47.4 - 49.0 & 1.87$\times$<2.72 (0.48) & 20 (5) & 93 (7)  & 70 (7) & 40 (10) & 36 & Maser\\
\hline

G35.03+0.35\_a & 18:54:00.554 (0.001) & +02:01:17.75 (0.07) & 54.8 & 54.4 - 55.2 & <3.25 & -- & 19 (2) & 48 (1) & 18 (2) & 3.6 & Maser$^{*}$\\
G35.03+0.35\_b & 18:54:01.058 (0.005) & +02:01:16.69 (0.01) & 52.8 & 52.0 - 53.6 & <1.63* & 23 (4) & 252 (3) & 130 (4) & 15 (5) & 186 & Maser\\
\hline

G45.47+0.05\_a & 19:14:25.671 (0.007) & +11:09:25.86 (0.10) & 64.4 & 63.2 - 64.8 & 1.61$\times$<0.90 (0.37) & 11 (2) & 18 (4) & 17 (2) & 23 (5) & 24 & Thermal\\
G45.47+0.05\_b & 19:14:25.679 (0.001) & +11:09:25.71 (0.02) & 66.4 & 65.2 - 67.2 & <0.90 & 13 (2) & 21 (1) & 41 (1) & 39 (2) & 51 & Maser$^{*}$

\enddata
\tablenotetext{a}{For sources for which the fitted size is poorly constrained (fitted size   $<2\times$ the statistical uncertainty from {\tt imfit}) and the S/N$<$50, we report the geometric mean of the synthesized beam as an upper limit on the source size (indicated with "<").  For sources for which the fitted size is poorly constrained (fitted size $<2\times$ the statistical uncertainty from {\tt imfit}) and the S/N$>$50, we report half the geometric mean of the synthesized beam as an upper limit on the source size (indicated with "<" and a * after the value.)}
\tablenotetext{b}{Dashes indicate non-detections (no emission above 4$\sigma$). The upper limits in these cases should be taken as 4 times the line rms from Table \ref{observationsTable}.}
\tablenotetext{c}{Velocity of the peak emission in the 5$_2$ line.}
\tablenotetext{d}{Velocity range over which emission was detected at the $\ge$4$\sigma$ level and {\tt imfit} was run.}
\tablenotetext{e}{Classification of emission as "thermal" or "maser", as described in \S~\ref{maser_v_thermal}.  Sources denoted ``maser$^{*}$'' are maser (non-thermal) emission, but their maximum fitted flux density occurs in a line higher than 5$_2$.}
\label{detailed}
\end{deluxetable*}
\end{turnpage}

Of the four observed \methanol\/ transitions, the 5$_2$ transition (see Table~\ref{lines}) is by far the most prevalent. Indeed, we detected this transition toward 16 of the 20 EGOs in the sample. In order to quantify the properties of the \methanol\/ emission, which is mostly very compact, we used the {\tt imfit} task in CASA to fit 2-dimensional Gaussians to each distinct emission component channel-by-channel. We limited the fitting to regions with emission $>4\sigma$, where $\sigma$ was measured locally to accurately assess the variable rms noise due to dynamic range limitations. The resulting fitted parameters for the position, velocity of peak emission, velocity range of emission, size, and flux density are given in Table~\ref{detailed}. The position, peak velocity, fitted flux density, and fitted size for each distinct spatial component are taken from the channel with the highest flux density for the 5$_2$ transition; only the fitted flux density in the peak channel is given for the other three transitions. The distinct emission regions are named by their galactic EGO name followed by a letter of the alphabet in order of increasing RA.

\subsubsection{Distinguishing maser and thermal emission}
\label{maser_v_thermal}

After fitting, we examined the properties of each distinct emission component and classified it as either maser or thermal emission. Ideally, we would use the line brightness temperature $T_B$ to discriminate between maser and thermal emission (i.e. $T_B$ exceeding a realistic thermal molecular gas temperature must be maser emission). For example, there are seven \methanol\/ components with 5$_2$ $T_B$ in excess of 1000~K, with a maximum of $10^5$~K for G08.67-0.35\_h (see Table~\ref{detailed}) that are clearly due to non-thermal emission. However, due to the relatively poor angular resolution of some of the data, especially those observed only in the D-configuration (see Table~\ref{observationsTable}), the current $T_B$ lower limits are not always constraining. This is particularly problematic because EGOs are thought to harbor massive star formation, and relatively warm thermal gas (few 100~K) is a natural consequence. Indeed, some of the observed EGOs are known to harbor hot core line emission with gas temperatures as high as a few 100~K \citep[see e.g.][]{Cyganowski2011a,Brogan11,C_g18,Cyganowski2014,Ilee2016}. In a few cases the 25~GHz \methanol\/ emission observed with the VLA is clearly thermal in origin, as evidenced by  spectral breadth (several \kms\/) and/or large fitted emission size (i.e. significantly larger than the beam). 

\begin{figure*}
    \figurenum{3}
    \includegraphics[width=0.9\textwidth]{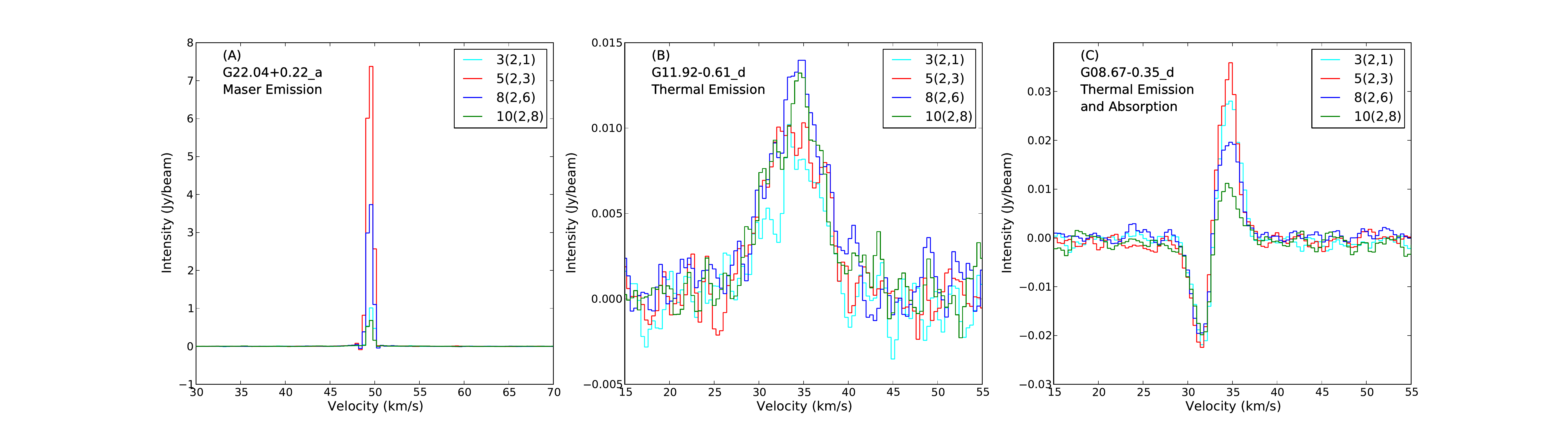}
    \caption{Example spectra for the two categories of emission (maser and thermal), as well as a spectrum showing both thermal absorption and emission. All three sets of spectra span 40 \kms\/.}
    \label{spectra}
\end{figure*}

To distinguish the emission mechanism for modest $T_B$ cases, we used two separate methods of analysis. The first uses the integer channel width (number of consecutive channels with emission $\geq$4$\sigma$ at the location of interest) and fitted angular size as discriminators. Spectrally broad ($\geq$4$\sigma$ in $\geq$5 channels, 2.0 \kms) emission with a large spatial extent (i.e. significantly spatially resolved fits) we classify as thermal emission. Emission that is spectrally narrow ($\geq$4$\sigma$ in $\leq$4 channels, 1.6 \kms) and spatially consistent with an unresolved point source we consider a candidate for maser emission. Within the category ``maser,'' there are two subcategories. Emission spots classified as ``maser'' are candidate maser emission, and their highest flux density is in the 5$_2$ transition. Emission spots classified as "maser$^{*}$" are likely to arise from non-thermal emission, but their highest flux density is in a transition other than 5$_2$ (usually 8$_2$). Figure~\ref{spectra} shows examples of maser and thermal spectra from our data.

The second method consisted of comparing our observed line ratios to line ratios produced by purely thermal, optically thin LTE emission. We numerically simulated line ratios for the optically-thin LTE case, and plotted these ratios (3$_2$:5$_2$, 8$_2$:5$_2$, and 10$_2$:5$_2$) for T=0 K to T=300 K. We then compared the observed line ratios for each fitted emission site to the simulated ratios. For lines with non-detections, we used an upper limit of 5$\sigma$ in the ratio, where $\sigma$ is the line rms from Table \ref{observationsTable}. Emission sites with line ratios inconsistent with LTE are candidates for maser emission. Ratios that match the simulated LTE emission indicate candidate thermal emission. 

Our findings with the second method largely matched our classifications from the first method. We found only two exceptions to our original classifications: G12.91-0.03\_b was classified as a maser, but its line ratios are consistent with optically thin LTE emission at lower temperatures ($\leq$ 40 K).  However, the T$_B$ for the 5$_2$ emission is 82 K, far warmer than the temperature required to produce optically thin thermal emission. We therefore consider G12.91-0.03\_b to be a maser. G22.04+0.22\_b was classified as thermal emission based on its large fitted size, but its line ratios are potentially more consistent with maser emission. However, G22.04+0.22\_b is in close proximity to G22.04+0.22\_a ($\sim1\farcs7$, and 2 channels), the strongest maser detected in our sample (7650 mJy in 5$_2$). The very strong emission from G22.04+0.22\_a causes ``ringing'' in the surrounding channels, including those in which G22.04+0.22\_b lies, so it is possible that the fitted flux densities of the 5$_2$ and 8$_2$ lines are skewed by this effect. It is also worth noting that G22.04+0.22\_b would be the only 25~GHz maser in the sample without a 44~GHz counterpart, and that the 25~GHz \methanol\/ emission is coincident with weak 1.3~cm emission (Table~\ref{cm_fluxes}), increasing the chance of warm thermal gas at this location. Thus, we consider G22.04+0.22\_b  as most likely ``thermal.''

\subsubsection{Detection Rates}

\begin{deluxetable*}{lccccccccc}[!hbt]
\tablecaption{25 GHz CH$_3$OH emission statistics by target}
\tablecolumns{10}
\tabletypesize{\footnotesize}
\tablehead{
\colhead{Source} & \colhead{25~GHz} & \multicolumn{4}{c}{Total 25~GHz Detections} & \colhead{Maser} & \colhead{Thermal} & \colhead{1.3 cm EGO}\\
\colhead{Name} & \colhead{Emission} & \colhead{3$_2$} & \colhead{5$_2$} & \colhead{8$_2$} & \colhead{10$_2$} & \colhead{Emission} & \colhead{Emission} & \colhead{Continuum}
}
\startdata
G08.67-0.35 & Y & 4 & 9 & 6 & 2 & 8 & 1 & Y\\
G10.29-0.13 & Y & 1 & 2 & 2 & 0 & 2 & 0 & Y\\
G10.34-0.14 & Y & 3 & 3 & 2 & 0 & 2 & 1 & N\\
G11.92-0.61 & Y & 4 & 5 & 2 & 1 & 4 & 1 & Y\\
G12.68-0.18 & Y & 1 & 2 & 2 & 2 & 1 & 1 & Y\\
G12.91-0.03 & Y & 1 & 2 & 1 & 0 & 2 & 0 & Y\\
G14.33-0.64 & Y & 3 & 4 & 3 & 2 & 4 & 0 & Y\\
G14.63-0.58 & Y & 1 & 1 & 0 & 0 & 0 & 1 & Y\\
G16.59-0.05 & Y & 2 & 2 & 2 & 2 & 1 & 1 & Y\\
G18.67+0.03 & Y & 0 & 1 & 1 & 0 & 0 & 1 & Y\\
G18.89-0.47 & N & 0 & 0 & 0 & 0 & 0 & 0 & N\\
G19.36-0.03 & Y & 3 & 4 & 2 & 2 & 4 & 0 & Y\\
G22.04+0.22 & Y & 3 & 3 & 3 & 2 & 2 & 1 & Y\\
G24.94+0.07 & Y & 0 & 1 & 0 & 0 & 0 & 1 & Y\\
G25.27-0.43 & N & 0 & 0 & 0 & 0 & 0 & 0 & N\\
G28.28-0.36 & Y & 1 & 1 & 1 & 1 & 1 & 0 & N\\
G28.83-0.25 & N & 0 & 0 & 0 & 0 & 0 & 0 & Y\\
G35.03+0.35 & Y & 1 & 2 & 2 & 2 & 2 & 0 & Y\\
G45.47+0.05 & Y & 2 & 2 & 2 & 2 & 1 & 1 & Y\\
G49.27-0.34 & N & 0 & 0 & 0 & 0 & 0 & 0 & Y\\
\enddata
\label{summary}
\end{deluxetable*}

Table \ref{summary} presents the number of 25~GHz \methanol\/ detections in each transition by source, as well as the total number of maser and thermal emission spots for each EGO, and whether 1.3~cm continuum emission is detected associated with the EGO (within or touching the boundary of the extended 4.5 \mum\/ emission, \S~\ref{cmresults}). Emission from different transitions is co-spatial so, e.g., co-spatial emission in the 3$_2$, 5$_2$, and 8$_2$ lines would be counted as a single maser spot in column 7.
We detected 25~GHz \methanol\/ emission above the 4$\sigma$ level in 16 of our 20 targets, for an overall detection rate of 80$\%$. For the 25~GHz Class I \methanol\/ masers, we found that the emission was strongest in the 5$_2$ transition (see Table \ref{stats}), but that the 8$_2$ transition was not significantly weaker than the 5$_2$ transition in general (median flux densities of 58.5 mJy and 52.0 mJy and median S/N ratios of 18.6 and 13.2 for 5$_2$ and 8$_2$, respectively). 
In total, we fit 44 sites of \methanol\/ emission. Of these 44 sites, we classified ten as being purely thermal emission. Of the remaining 34 sites, we classified 30 as ``maser'' and 4 as ``maser$^{*}$'' (as defined in \S~\ref{maser_v_thermal}).  Detection rates and flux density statistics by transition and emission type are summarized in Table \ref{stats}.

\begin{deluxetable*}{lccccc}[hbt]
\tablecaption{25 GHz \methanol\/ Emission Statistics by Transition}
\tablecolumns{6}
\tabletypesize{\footnotesize}
\tablehead{
\colhead{Transition} & \colhead{Detection} & \colhead{Median Flux} & \colhead{Mean Flux} & \colhead{Median } & \colhead{Mean}\\
 & \colhead{Rate} & \colhead{Density (mJy)} & \colhead{Density (mJy)} & \colhead{ S$_{peak}$/N} & \colhead{ S$_{peak}$/N}\\
\hline
\multicolumn{6}{c}{Maser Detection Statistics}
}
\startdata
3(2,1)-3(1,2) & 60\% (12/20) & 30.0 & 261.7 & 7.8 & 47.4\\
5(2,3)-5(1,4) & 80\% (13/20) & 58.5 & 613.2 & 18.6 & 120.7\\
8(2,6)-8(1,7) & 80\% (13/20) & 52.0 & 327.1 & 13.2 & 58.3\\
10(2,8)-10(1,9) & 45\% (9/20) & 33.5 & 98.3 & 9.7 & 19.0\\
\cutinhead{Thermal Detection Statistics}
3(2,1)-3(1,2) & 40\% (8/20) & 28.5 & 48.4 & 8.9 & 21.9\\
5(2,3)-5(1,4) & 50\% (10/20) & 34.5 & 55.4 & 10.6 & 21.5\\
8(2,6)-8(1,7) & 40\% (8/20) & 38.0 & 48.8 & 12.4 & 17.6\\
10(2,8)-10(1,9) & 30\% (6/20) & 41.0 & 39.5 & 14.4 & 17.4
\enddata
\label{stats}
\tablecomments{Columns 5 and 6 list the mean and median signal-to-noise (S$_{peak}$/N) for detections in each transition, where S$_{peak}$/N is calculated for each maser using the line rms from Table \ref{observationsTable}.}
\end{deluxetable*}

Of the sites identified as thermal emission, one has emission only in the 5$_2$ line, one has emission only in the 3$_2$ and 5$_2$ lines, one has emission only in the 5$_2$ and 8$_2$ lines, and one has emission in the 3$_2$, 5$_2$, and 8$_2$ lines (Table~\ref{detailed}). The remaining six sites have thermal emission in all four lines. For the source with thermal emission in 5$_2$ and 8$_2$ only (G18.67+0.03\_a), we did identify weak thermal emission in the 3$_2$ line, but it is not above the 4$\sigma$ level and so is not included in Table~\ref{detailed}. 

The 34 emission sites classified as ``maser'' or ``maser$^{*}$'' have the following properties: 10 have emission above the 4$\sigma$ level (where $\sigma$ is the line rms from Table~\ref{observationsTable}) in all four transitions (3$_2$, 5$_2$, 8$_2$, 10$_2$), including three of the four spots classified as ``maser$^{*}$''; eight have emission above the 4$\sigma$ level in the 3$_2$, 5$_2$, and 8$_2$ lines only; two have emission above the 4$\sigma$ level in the 5$_2$, 8$_2$, and 10$_2$ lines only; four have emission above the 4$\sigma$ level in the 3$_2$ and 5$_2$ lines only; three have emission above the 4$\sigma$ level in the 5$_2$ and 8$_2$ lines only; and seven have emission only in the 5$_2$ line. For this last group, the median and mean flux densities are 27.0~mJy and 35.9~mJy, respectively. These values are lower than the median and mean flux densities for the 5$_2$ population as a whole (see Table~\ref{stats}), but these masers are not the weakest masers in the overall population.

\subsection{Notes on Individual Sources}\label{individual}
The following subsections discuss each of the target EGOs for which we detect either 25~GHz \methanol\/ or EGO-associated 1.3~cm continuum emission in greater detail, including notes on relevant high-resolution observations reported in the literature. 

\paragraph{G08.67$-$0.35}
 We detect thermal 25~GHz \methanol\/ in both emission and absorption (G08.67-0.35\_d; see Fig.~\ref{spectra}) and resolved 1.3~cm continuum emission coincident with the known UC\HII\/ region G8.67-0.36 \citep{Wood1989}. The northern edge of the UC\HII\/ region is coincident with the southern end of the extended 4.5 \mum\/ emission of the G08.67$-$0.35 EGO (Fig.~\ref{3color}).  It is currently unclear whether the extended 4.5~\mum\/ emission arises from an outflow associated with the UC\HII\/ region or from an outflow driven by a less-evolved source that is (as yet) undetected in 1.3~cm continuum. 
 Four of the 25~GHz \methanol\/ masers are coincident with the extended 4.5 \mum\/ emission, while three more lie in an arc 5-8$\arcsec$ west and north of it. 
The thermal 25~GHz \methanol\/ emission is coincident with the UC\HII\/ region and with the Class II 6.7~GHz \methanol\/ maser emission (G8.669$-$0.356) reported by Caswell (2009).

\paragraph{G10.29$-$0.13}
Like \citet{Cyganowski2011b}, we detect 1.3~cm continuum emission associated with the MIR-bright (MIPSGAL-saturated) source directly to the east of the EGO.  Both the MIR-bright source and the EGO lie on the edge of the W31 \HII\/ region G10.32$-$00.15 (\citealt{Westerhout1958}, see also discussion in \citealt{Cyganowski2011b}). In our image, the 1.3~cm continuum emission partially overlaps the extended 4.5 \mum\/ emission of the EGO and so is included in Tables~\ref{cm_fluxes} and \ref{summary} as an EGO-associated 1.3~cm source (\S~\ref{cmresults}).  We note, however, that the morphology of the 1.3~cm continuum differs markedly from that of the extended 4.5 \mum\/ emission, and it is unclear if the two are really associated. 
Indeed, \citet{Cyganowski2011b} deem their 1.3 and 3.6~cm detections to be unrelated (at the higher resolution and poorer sensitivity of those data there is no overlap in the centimeter continuum and extended 4.5~\mum\/ emission).
Higher angular resolution and sensitivity continuum observations are needed to verify that there is cm-$\lambda$ emission directly associated with the EGO.

Unlike most EGOs, G10.29$-$0.13 lacks a discrete  24~\mum\/ counterpart, though it is associated with a 6.7 GHz Class II \methanol\/ maser \citep{Cyganowski2009}. 
We do not detect any 25~GHz \methanol\/ emission coincident with this EGO (Fig.~\ref{3color}).
 We do detect two 25~GHz \methanol\/ masers in the field, located $\sim 25\arcsec$ NW and SE of the EGO, respectively, that appear to be distributed along the same line as the 44~GHz Class I \methanol\/ masers reported in \citet{Cyganowski2009}. Neither 25 GHz maser is associated with 1.3~cm continuum or extended 4.5 \mum\/ emission.

\paragraph{G10.34$-$0.14}
G10.34-0.14 is also located on the edge of the W31 \HII\/ region G10.32$-$00.15 (also see \S {\it G10.29$-$0.13}). We do not detect 1.3~cm continuum emission associated with the EGO.  Within the VLA FOV, we detect extended 1.3~cm continuum emission from the nearby MIR-bright \HII\/ region and a weak, unresolved 1.3~cm source $\sim 18\arcsec$ west of the EGO (Fig.~\ref{3color}; not included in Table~\ref{cm_fluxes}). 

We detect 25~GHz thermal \methanol\/ emission coincident with the EGO and with Class II 6.7~GHz \methanol\/ maser emission \citep{Cyganowski2009}. The thermal \methanol\/ emission is fairly extended, with a T$_b$ of only 6 K (Table~\ref{detailed}), and is coincident with the southern edge of a N-S elongated region of 24 \mum\/ emission.  This MIR morphology may indicate the presence of two blended 24 \mum\/ sources. 
One 25~GHz \methanol\/ maser is detected at the NW end of the extended 4.5 \mum\/ emission, coincident with a 44~GHz Class I \methanol\/ maser reported by \citet{Cyganowski2009}. The other 25~GHz maser is SE of the EGO, coincident with a separate patch of extended 4.5 \mum\/ emission and 44~GHz \methanol\/ masers \citep{Cyganowski2009}.

\paragraph{G11.92$-$0.61}
We detect three 1.3~cm continuum sources, two coincident with the EGO (CM1 and CM2, Table~\ref{cm_fluxes}) and one at the SE edge of the extended 4.5~\mum\/ emission (Fig.~\ref{3color}).  The strongest 1.3~cm source, CM1 \citep[previously reported by ][]{Cyganowski2011b,Cyganowski2014,Moscadelli2016,Ilee2016}, is coincident with the millimeter dust source and massive disk candidate MM1 \citep{Cyganowski2011a,Ilee2016} and with 6.7~GHz \methanol\/ maser emission \citep{Cyganowski2009}.  Based on modelling the centimeter-submillimeter spectral energy distribution (SED) of MM1, \citet{Ilee2016} argue that its cm-$\lambda$ emission is attributable to free-free emission from a gravitationally trapped hypercompact (HC) \HII\/ region, with a possible contribution from a compact ionized jet \citep[see also][]{Moscadelli2016}.  We detect 25~GHz thermal \methanol\/ emission coincident with CM1, consistent with the classification of MM1 as a hot core based on inteferometric (sub)millimeter line observations \citep{Cyganowski2011a,Cyganowski2014,Ilee2016}.

Located $\sim3\arcsec$ north of CM1, CM2 is also coincident with 6.7~GHz Class II \methanol\/ maser emission \citep{Cyganowski2009}.  This weak, unresolved 1.3 cm source ($\sim$0.3 mJy beam$^{-1}$, Table~\ref{cm_fluxes}) is the cm-$\lambda$ counterpart of the millimeter source MM3$-$C1 \citep{Cyganowski2011a,Cyganowski2017}, and was detected at 0.9 and 3 cm with the VLA by \citet{Cyganowski2017}. 
The third 1.3 cm source is $\sim13\arcsec$ SE of CM1, toward the edge of the extended 4.5 \mum\/ emission. This centimeter detection is weak and unresolved, and is not associated with compact 24 \mum\/ or (sub)millimeter emission. It is located within 1$\arcsec$ of a 44~GHz Class I \methanol\/ maser \citep{Cyganowski2009}, but the relationship between the maser and the centimeter emission is unclear.

We detect four 25~GHz \methanol\/ masers in this target, all of which have 44~GHz \methanol\/ maser counterparts. These masers are all located toward the edges of the extended 4.5 \mum\/ emission, and are distinctly separated (> 5$\arcsec$) from the centimeter emission and MIPSGAL 24 \mum\/ peak.

\paragraph{G12.68$-$0.18}
We detect a compact 1.3~cm continuum source coincident with the EGO, as well as extended ($> 15\arcsec$) 1.3~cm emission coincident with similarly extended MIPSGAL 24 \mum\/ emission. The EGO-related centimeter source, which we denote CM1, was studied by \citet{Moscadelli2016} with the VLA.  Based on their high-resolution multiwavelength observations (resolution $\lesssim$0\farcs3 at 4.8, 2.3, and 1.4 cm), \citet{Moscadelli2016} suggest that the cm-$\lambda$ continuum emission arises from an ionized jet.        

CM1 is coincident with both thermal and masing 25~GHz \methanol\/ emission (Fig.~\ref{3color}), and with a 6.7 GHz Class II \methanol\/ maser (\citealt{Caswell2009}, see also discussion in \citealt{Moscadelli2016}).
The 25~GHz thermal emission has a brightness temperature of 90 K, suggestive of warm gas on small size scales.  This is consistent with the results from Submillimeter Array (SMA) observations of the W33 complex by \citet{Immer2014}, who find that the millimeter continuum counterpart to CM1 (W33B; see their Fig.~6) is a hot core rich in nitrogen-bearing species, with gas temperatures of $\sim$220-350 K.

\paragraph{G12.91$-$0.03}
We detect one weak, unresolved 1.3~cm continuum source and two 25~GHz \methanol\/ masers in the VLA FOV.  The 1.3~cm source, which we denote CM1, is located at the NE edge of the extended 4.5 \mum\/ emission (Fig.~\ref{3color}).
One of the 25~GHz masers is coincident with the extended 4.5~\mum\/ emission, 24 \mum\/ emission, and the 6.7~GHz Class II \methanol\/ maser G12.904$-$-0.031 \citep{Green2010}. The second 25~GHz maser is located just beyond the western edge of the extended 4.5~\mum\/ emission, toward the edge of a more evolved, 8~\mum-bright region.

\paragraph{G14.33$-$0.64}
This EGO is located $\sim 15\arcsec$ SE of the bright far-infrared source IRAS 18159-1648 \citep{Jaffe82}, within a ridge of ammonia emission \citep[VLA observations by][]{Lu2014}.  We detect marginally-resolved 1.3~cm continuum emission coincident with the EGO (denoted CM1) and also the IRAS source (Fig.~\ref{3color}).  The morphology of CM1 is consistent with two unresolved cm-$\lambda$ continuum sources. Both components have emission above 6$\sigma$, but satisfactory two-component fits could not be achieved with the current data.

We detect four 25~GHz \methanol\/ masers and no thermal \methanol\/ emission in this source.  The 25~GHz \methanol\/ emission is, however, confused both spatially and spectrally. 
Consequently, there may be additional weak \methanol\/ emission present that could not be separated in the current data.  Two of the 25~GHz masers are located just within the 4$\sigma$ contour of the 1.3~cm continuum emission; two are located north of the centimeter source.

\paragraph{G14.63$-$0.58}
We detect two 1.3~cm continuum sources, both coincident with extended 4.5 \mum\/ emission (Fig.~\ref{3color}).  The brighter centimeter source, CM1, is also coincident with compact 24~\mum\/ emission and with the 6.7~GHz \methanol\/ maser G14.631$-$0.577, reported by \citet{Green2010}.  Thermal 25 GHz \methanol\/ emission is detected towards CM2, the weaker centimeter continuum source.
The brightness temperature of this thermal \methanol\/ emission is only T$_b$ = 2.9 K, due to the extended nature of the emission and consequent large fitted size; our optically-thin calculation method gives a temperature range of T = 10 - 40 K.  No 25~GHz \methanol\/ masers are detected in the field.

\paragraph{G16.59$-$0.05}
This EGO is adjacent to IRAS 18182$-$1433 \citep[nominal separation $\sim$19\arcsec;][]{egocat}; unusually among our sample, its cm-$\lambda$ continuum emission has been well-studied, primarily by authors targeting the IRAS source \citep[e.g.][]{Zapata2006,Sanna2010,Hofner2011,Moscadelli2013,Moscadelli2016,Rosero16}. 
We detect a single compact 1.3~cm continuum source (CM1), which is coincident with the EGO and with a local peak in the 24 \mum\/ emission (Fig.~\ref{3color}).  CM1 corresponds to the brightest of the five components (18182-1433~C) detected at both 1.3~cm and 6~cm in deep VLA observations of this field by \citet{Rosero16}, who measure a spectral index of $+0.8\pm0.1$ for this object.  The compact 1.3~cm source is coincident with 6.7~GHz \methanol\/ maser emission \citep[e.g.][]{Green2010,Sanna2010,Moscadelli2013}.  At longer wavelengths, the continuum emission is elongated E-W, with a sizescale of $\sim$4\arcsec\/ at 6~cm \citep{Moscadelli2013,Moscadelli2016}.  The orientation of this elongation, which is interpreted as an ionized jet \citep[e.g.][]{Moscadelli2013,Moscadelli2016}, is notably similar to that of the extended 4.5~\mum\/ emission of the EGO, which is elongated E-W on larger scales ($\sim$10-15\arcsec).

We detect thermal 25~GHz \methanol\/ emission that is coincident with the EGO and CM1, and with the 6.7~GHz \methanol\/ maser G16.585$-$0.051 reported by \citet{Green2010}.  The detection of thermal \methanol\/ at 25~GHz is consistent with the identification of this source as a hot core by \citet{Beuther2006} using SMA observations and by \citet{Lu2014} using VLA observations.  We also detect one 25~GHz maser, located $<1\arcsec$ S-SE of the thermal \methanol\/ emission.

\paragraph{G18.67$+$0.03}
We detect two sources of 1.3~cm continuum emission in the VLA FOV (Fig.~\ref{3color}).  CM1 is weak, unresolved, and coincident with the EGO, a compact MIPSGAL 24 \mum\/ source, and 6.7~GHz Class II \methanol\/ maser emission (\citealt{Cyganowski2009,Green2010}, see also \citealt{C_g18}).  The measured peak intensity of CM1 (0.18 \mjb, Table~\ref{cm_fluxes}) is consistent with this source being undetected by \citet{Cyganowski2011b} (4$\sigma$ upper limit of 0.94 \mjb\/ at 1.3~cm). 
We also detect strong, resolved 1.3~cm continuum emission $\sim13''$ W of the EGO, from the source designated ``UCHII'' by \citet{C_g18} \citep[F G18.67+0.03-CM1 in][]{Cyganowski2011b}.     

The only 25~GHz \methanol\/ detection in the field is thermal, and is coincident with CM1 and 6.7 GHz \methanol\/ maser emission.  Unusually, \methanol\/ emission is detected in the 5$_2$ and 8$_2$ lines, but not 3$_2$.  The brightness temperature T$_b$ = 15 K, with the optically-thin LTE calculations suggesting a temperature range of 40 - 50 K.
The detection of thermal \methanol\/ at 25 GHz is consistent with the presence of strong hot core molecular line emission from the EGO in SMA observations by \citet{C_g18}, who find a CH$_3$CN temperature of 175~K.  Notably, though both the EGO and the UC HII region to the west are associated with 44 GHz Class I \methanol\/ masers \citep{Cyganowski2009,C_g18}, no 25 GHz \methanol\/ maser emission is detected within the VLA field of view.

\paragraph{G18.89$-$0.47}
No 1.3~cm continuum or 25~GHz \methanol\/ line emission was detected toward this EGO.

\paragraph{G19.36$-$0.03}
We detect three sources of 1.3~cm continuum emission in the field, two of which (CM1 and CM2) are associated with MIPSGAL 24 \mum\/ emission (Fig.~\ref{3color}).  The 24 \mum\/ emission is elongated (NW-SE); its morphology suggests at least two components, separated by $\sim6\arcsec$.  The weak 1.3~cm source CM1 (1.3 mJy, Table~\ref{cm_fluxes}) is associated with the NW 24 \mum\/ component and coincides with the previously detected 3.6~cm source denoted F-CM1 by \citet{Cyganowski2011b}.  

CM2 is a new cm-$\lambda$ detection and is coincident with the EGO and with Class II 6.7~GHz \methanol\/ maser emission \citep{Cyganowski2009}.  CM2 is unresolved and weak (0.40 mJy, Table~\ref{cm_fluxes}), consistent with the previous $4\sigma$ upper limit of 1~mJy at 1.3~cm \citep{Cyganowski2011b}.  The third 1.3 cm continuum source lies partially off the southern edge of the field shown in Figure \ref{3color}, and was also detected by \citet{Cyganowski2011b} at 3.6~cm (their F-CM2).  

We detect four 25~GHz Class I \methanol\/ masers in this field, all of which have 44~GHz counterparts. The masers lie approximately along a line connecting the northern edge of CM1 and an arc of 44~GHz \methanol\/ masers to the SW, including one that is near CM2.

\paragraph{G22.04$+$0.22}
We detect two 1.3~cm continuum sources in the field.  CM1 is a weak (0.6 \mjb, Table~\ref{cm_fluxes}), unresolved source that is coincident with the EGO, compact 24 \mum\/ emission, and 6.7~GHz \methanol\/ masers \citep{Cyganowski2009}. A second centimeter continuum source is detected $\sim 15\arcsec$ NW of the EGO, and does not appear to be associated. Neither centimeter continuum source was detected by \citet{Cyganowski2011b}, consistent with their 4$\sigma$ upper limit of 1~mJy at 1.3~cm. 

We detect thermal 25~GHz \methanol\/ emission coincident with CM1 and the 6.7~GHz masers (Fig.~\ref{3color}).
We also detect two 25~GHz \methanol\/ masers, one of which is $<$2$\arcsec$ SW of CM1 and its thermal 25~GHz \methanol\/ emission.  The second 25 GHz maser is coincident with a line of 44~GHz \methanol\/ masers that extends to the south of the EGO \citep{Cyganowski2009}.

\paragraph{G24.94$+$0.07}
We detect weak 1.3~cm continuum emission coincident with this EGO (0.85 \mjb, Table~\ref{cm_fluxes}).  The 1.3~cm source, CM1, is coincident with compact MIPSGAL 24 \mum\/ emission and 6.7 GHz \methanol\/ masers \citep{Cyganowski2009}.  CM1 was detected by \citet{Cyganowski2011b} at 3.6~cm (peak intensity 0.53$\pm$0.04 \mjb); our 1.3~cm detection is consistent with their 4$\sigma$ upper limit of 1.0 mJy  at 1.3~cm.  Unfortunately the mismatch in beam size between the 3.6 and 1.3~cm detections precludes analysis of the cm-wavelength SED.
The only 25~GHz \methanol\/ emission detected in this source is thermal emission associated with CM1 (Fig.~\ref{3color}).


\paragraph{G25.27$-$0.43}
No 1.3~cm continuum emission is detected towards this EGO.  An evolved \HII\/ region located $\sim30\arcsec$ SE of the EGO is detected in 1.3~cm continuum; this source was detected by \citet{Cyganowski2011b} at 3.6~cm, who designated it F-CM2. 
No 25~GHz \methanol\/ emission was detected within the VLA field of view. 

\paragraph{G28.28$-$0.36}
The only 1.3~cm continuum emission detected in this field is strong, resolved emission associated with the well-known core-halo UC \HII\/ region G28.288-0.364 \citep[e.g.][]{Kurtz1994}, $\sim$20$\arcsec$ E-NE of the EGO (Fig.~\ref{3color}).  This UC \HII\/ region was detected at 3.6~cm and 1.3~cm by \citet{Cyganowski2011b}, who designated it F-CM1.  Notably, we do not detect 1.3~cm continuum emission from the \citet{Cyganowski2011b} 3.6~cm source CM1, which is coincident with the EGO and $\sim$1\farcs2 NE of a 6.7 GHz \methanol\/ maser \citep{Cyganowski2009,Cyganowski2011b}.  The relatively high rms noise of the new 1.3~cm VLA image for this source (the third-highest of our sample, Table~\ref{observationsTable}) means that our 1.3~cm limit is only a factor of $\sim$1.4 improvement over that of \citet{Cyganowski2011b}, and the mismatch in beam size precludes combining the two datasets to better constrain CM1's cm-$\lambda$ spectral index.
Only one 25~GHz \methanol\/ maser is detected within the VLA field of view; this maser is $\sim7\arcsec$ north of F-CM1 and is coincident with a 44~GHz \methanol\/ maser reported by \citet{Cyganowski2009}.

\paragraph{G28.83$-$0.25}

We detect two weak 1.3~cm continuum sources ($<$0.4 \mjb, Table~\ref{cm_fluxes}) coincident with this EGO. 
The eastern centimeter source, CM2, is also coincident with compact MIPSGAL 24 \mum\/ emission and 6.7 GHz \methanol\/ masers \citep{Cyganowski2009}.  The 1.3~cm emission from this source  
is spatially extended E-W, with a morphology consistent with multiple unresolved or marginally-resolved sources (Fig.~\ref{3color}).  The western centimeter source, CM1, is unresolved and located $\sim$4$\arcsec$ west of CM2.
Both CM1 and CM2 were detected by \citet{Cyganowski2011b} at 3.6~cm (but not at 1.3~cm; their 1.3~cm 4$\sigma$ upper limit 
was 0.92 mJy beam$^{-1}$).  
Interestingly, at 3.6~cm the western source (CM1 in both papers) is the brighter of the two, while at 1.3~cm CM2 is the brighter source \citep[Table~\ref{cm_fluxes} and][]{Cyganowski2011b}.  This reversal suggests that CM2 either has a steeper free-free SED or more contribution from dust than CM1.   
We do not detect any 25~GHz \methanol\/ emission above the 4$\sigma$ level for this EGO.  In at least three of the four \methanol\/ lines there is, however, very weak emission ($\sim 3\sigma$, so not included in our analysis), likely thermal, coincident with CM2.

\paragraph{G35.03$+$0.35}
The 1.3~cm VLA survey data for this source were presented in \citet{Brogan11}.
As shown in Figure~\ref{3color}, there is strong, spatially extended 1.3~cm continuum emission coincident with the EGO. 
\citet{Cyganowski2011b} resolved five distinct sources at 3.6~cm (denoted CM1..CM5).  The two strongest (CM1 and CM2) were also detected by these authors at 1.3~cm, and CM2 is associated with 6.7 GHz \methanol\/ masers \citep{Cyganowski2009,Cyganowski2011b,Surcis2015}.  
The morphology of the 1.3~cm continuum emission in our VLA image is consistent with multiple, unresolved centimeter sources, and is spatially coincident with CM1, CM2, CM4, and CM5 from \citet{Cyganowski2011b}.  As we could not obtain satisfactory multi-component fits to the current data, we report the combined total 1.3~cm flux density (18.5 mJy) as a single 1.3~cm source, which we denote CM1 (Table~\ref{cm_fluxes}). 

We detect one 25~GHz \methanol\/ maser coincident with the 1.3~cm continuum emission, and one at the eastern edge of the extended 4.5 \mum\/ emission, coincident with an arc of 44~GHz Class I \methanol\/ masers reported by \citet{Cyganowski2009}.  \citet{Brogan11} reported the 25~GHz \methanol\/ maser results from these VLA data, along with the detection of an NH$_3$ (3,3) maser coincident with the 44~GHz Class I \methanol\/ maser arc. 
Using H63$\alpha$ and H64$\alpha$ recombination lines, \citet{Brogan11} find a velocity of 55.8 \kms\/ for the free-free emission from CM1, in good agreement with the velocity of the coincident 25~GHz \methanol\/ maser. Unfortunately, the \citet{Cyganowski2009} observations at 44~GHz only extend up to $\sim$54.4 \kms, and so the velocity at which the maser G35.03+0.35\_a lies is not covered by the 44~GHz data.
However, the spectrum presented in \citet{Kang2015} suggests an upper limit of $\sim$1~Jy.

\paragraph{G45.47$+$0.05}
We detect strong, resolved 1.3~cm continuum emission coincident with the EGO and with the known UC \HII\/ region G45.47+0.05 \citep{Wood1989,Hofner1999}, classified by \citet{Wood1989} as ``irregular or multiply peaked.''
In the new 1.3~cm VLA image, CM1 appears elongated along a NW-SE axis, suggestive of a possible ionized jet (Fig.~\ref{3color}). The elongation direction of the cm-$\lambda$ emission matches that of the extended 4.5 \mum\/ emission.

We detect both thermal and masing 25~GHz \methanol\/ emission for this EGO, both coincident with a compact, southern component of CM1.  The thermal emission is fairly compact compared to other thermal \methanol\/ detections in the survey (1.61$\times$<0.90$\arcsec$), and has T$_b$ = 24 K. The optically-thin LTE calculations suggest a physical temperature of T $\sim$100 K.

\paragraph{G49.27$-$0.34}
We detect strong, resolved 1.3~cm continuum emission coincident with this EGO (Fig.~\ref{3color}).  This centimeter source, CM1, was detected by \citet{Cyganowski2011b} at 3.6~cm and 1.3~cm and by \citet{Mehringer1994} at 20~cm; \citet{Cyganowski2011b} found that its cm-$\lambda$ spectral index was consistent with optically-thin free-free emission. 
The 1.3~cm continuum emission from CM1 exhibits a roughly circular morphology and is coincident with both extended 4.5 \mum\/ and 24 \mum\/ emission \citep[see also discussion in][]{Cyganowski2011b}.
We do not detect a 1.3~cm counterpart to the weak, compact \citet{Cyganowski2011b} 3.6~cm source CM2 (0.61 \mjb\/ at 3.6~cm) at the $>$4$\sigma$ level \citep[4$\sigma$ limit of 0.28 \mjb\/ compared to 0.71 \mjb\/ in][]{Cyganowski2011b}. 
Neither thermal nor masing 25~GHz \methanol\/ emission is detected towards this EGO.

\section{Discussion}
\label{discussion}

\subsection{Spatial Distribution of 25~GHz CH$_3$OH Emission Compared to 4.5~\mum\/ Emission}
\label{spatial_cont}
The overall correlation between 25~GHz \methanol\/ emission and extended 4.5~\mum\/ emission is strong. 
In only two of our target regions is the 25~GHz emission entirely outside the boundaries of the extended 4.5~\mum\/ emission. In these cases (G10.29-0.13 and G28.28-0.36), the relationship between the 4.5~\mum\/ emission and the 25~GHz \methanol\/ masers is unclear.
In total, 25 of our 34 detected maser sites (74\%) are coincident with extended 4.5~\mum\/ emission. The nine masers 
that are not coincident with extended 4.5~\mum\/ emission are predominantly located near dark clouds (Fig.~\ref{3color}).  

Nine of the ten thermal \methanol\/ detections (90\%) are coincident with extended 4.5~\mum\/ emission (see Table \ref{summary}); of these, all but one (G14.63$-$0.58) is also coincident with strong 24~\mum\/ emission (Fig.~\ref{3color}). The exception is the thermal \methanol\/ emission in G08.67-0.35, which is not coincident with the extended 4.5~\mum\/ emission but is instead coincident with 1.3~cm continuum emission and the known \HII\/ region G8.67-0.36 \citep[][see also \S \ref{individual}]{Wood1989}.

\subsection{Spatial Distribution of 25~GHz CH$_3$OH Emission Compared to 1.3~cm Continuum Emission}
We find a strong correlation between the presence of thermal \methanol\/ emission at 25~GHz and the presence of 1.3~cm continuum emission.
Nine of our ten thermal emission detections (90\%) are coincident with a 1.3~cm detection (see Table \ref{summary}). The exception is G10.34-0.14, which shows one thermal emission site but has no detected 1.3~cm continuum emission coincident with the EGO.
However, there is strong 1.3~cm continuum emission to the south from the W31 star-forming complex, and this causes the G10.34-0.14 field to have one of the poorer continuum sensitivities due to dynamic range limitations. This may be limiting our ability to detect weak 1.3~cm emission towards the EGO in this case.

While sources that have thermal \methanol\/ almost always have 1.3~cm continuum emission, we find that the reverse is not true: of our 19 sources of 1.3~cm emission, only nine (47\%) have coincident thermal \methanol\/ emission. In addition, we also find only a weak correlation between the presence of 25~GHz \methanol\/ maser emission and the presence 1.3~cm continuum emission, with only only 8 masers (24\%) coincident with continuum (where we define ``coincident'' as being within the boundary of the 4$\sigma$ level of the 1.3~cm emission), while the other 26 masers lie outside the boundaries of any 1.3~cm continuum.

\subsection{Correlation Between 6.7~GHz CH$_3$OH Masers, 1.3~cm Continuum, and Thermal 25~GHz CH$_3$OH Emission}
Interestingly, 6.7~GHz Class II \methanol\/ masers do appear to be correlated with both 1.3~cm continuum and 25 GHz thermal \methanol\/ emission in our EGO sample. 
Nineteen of the 1.3~cm sources are in regions for which past 6.7~GHz data exist in the literature (the exception being G14.33$-$0.64\_a, Table~\ref{distances}). Twelve of these nineteen 1.3~cm sources (63\%) are coincident with 6.7~GHz masers (\S \ref{individual}); of these, eight are new detections at 1.3~cm, and four (G14.63\_CM1, G18.67\_CM1, G19.36\_CM2, and G22.04\_CM1) are, to our knowledge, new detections at any cm-$\lambda$. Conversely, twelve of the eighteen Class II 6.7 GHz \methanol\/ masers associated with our target EGOs (67\%) are coincident with 1.3~cm emission in the VLA images.  Of the six 6.7~GHz masers without cm-$\lambda$ detections, three are in regions (G10.29$-$0.13, G10.34$-$0.14, and G28.28$-$0.36) that have 2-6 times poorer sensitivity than the majority of the sample due to dynamic range limitations (Table~\ref{observationsTable}).      
The detection of weak cm-$\lambda$ continuum emission associated with Class II \methanol\/ masers is consistent with both phenomena tracing young, deeply embedded massive (proto)stars.  Similarly, thermal 25~GHz \methanol\/ emission in the VLA data may pinpoint hot core emission, which has been observed in association with 6.7~GHz masers in large-scale single-dish surveys \citep[e.g.][]{Purcell2006,Purcell2009}.
Of the ten sources of thermal 25~GHz \methanol\/ emission seen with the VLA, eight are coincident with 6.7~GHz masers. The exceptions are G14.63-0.58\_a, in which the thermal \methanol\/ emission is coincident with the weaker centimeter source CM2 but the 6.7~GHz masers are coincident with the stronger centimeter source CM1, and G45.47+0.05\_a, which has thermal \methanol\/ emission but no 6.7~GHz masers.  High-resolution observations in other hot-core tracers (e.g. with (sub)millimeter interferometers) will illuminate the nature of these objects.

\subsection{Detailed Comparison of 25~GHz and 44~GHz Class I \methanol\/ Masers}
\label{matching}

\begin{figure*}
    \figurenum{4}
    \centering
    \includegraphics[width=0.9\textwidth]{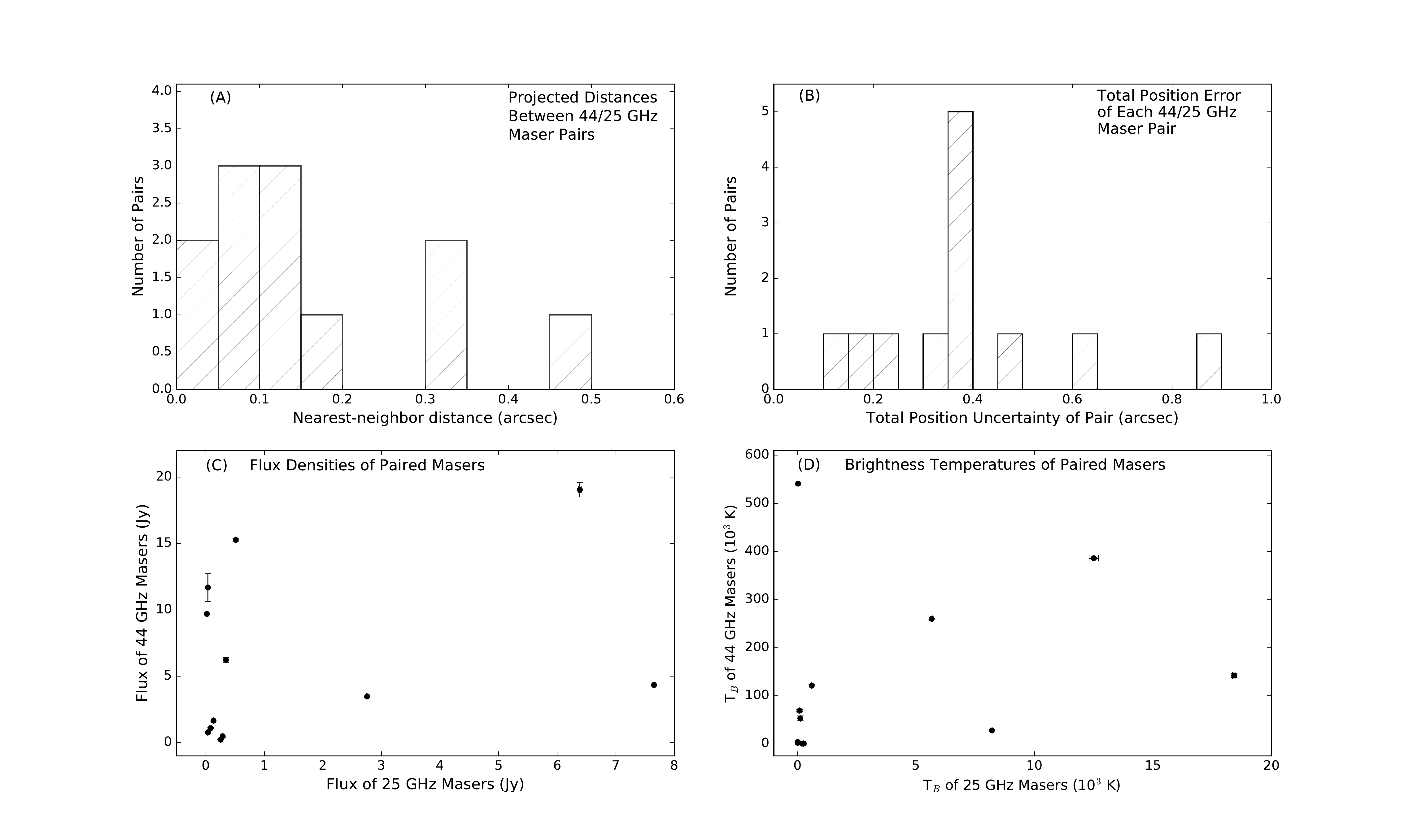}
    \caption{(a) Nearest-neighbor distances, (b) position uncertainties, (c) 44~GHz vs. 25~GHz fluxes, and (d) 44~GHz vs. 25~GHz brightness temperatures of paired masers.}
    \label{pairs}
\end{figure*}

Current models for the pumping of Class I \methanol\/ masers suggest that 
25 GHz masers are excited under an overlapping but narrower region of parameter space than 44 GHz masers \citep{Sobolev2007}. Thus, we do not expect 25 GHz masers to be present without a 44 GHz counterpart.  In order to test this hypothesis observationally, we compare our 25~GHz \methanol\/ maser properties with the 44~GHz \methanol\/ maser results from the VLA survey of \citet{Cyganowski2009}. Of the 13 EGO fields with 25~GHz  \methanol\/ maser detections, seven were also included in the \citet{Cyganowski2009} survey. 
A total of 17 individual 25~GHz \methanol\/ masers 
were detected toward the EGOs in common between the two surveys. Of these 17 masers, two (G10.29-0.13\_b and G35.03+0.35\_a) do not have complementary 44~GHz data because the spectral breadth of the 44 GHz observations did not cover the 25 GHz maser velocity. This leaves 15 masers with interferometric observations at both 25~GHz and 44~GHz. 
In order to compare the 25 and 44~GHz \methanol\/ masers, we first regridded the \citet{Cyganowski2009} 44~GHz image cubes (with a channel width of 0.17~\kms\/) to 0.4~\kms\/ channels to match the 25~GHz data. After testing, we elected not to convolve the 44~GHz image cubes to the poorer angular resolution of the 25~GHz data. The 44~GHz data have significantly higher angular resolution ($0\farcs5$ to $1\farcs0$) for most sources, and there are typically many more 44~GHz masers in a given region than 25~GHz masers (see Fig.~\ref{3color}).  Preserving the higher angular resolution of the 44 GHz data allows us to pinpoint which 44~GHz maser provided the closest positional match for each 25 GHz detection. 

For each of the fifteen 25 GHz masers with 44 GHz data, the properties of the 44~GHz maser in closest positional and kinematic proximity (typically matching to within one channel) were fit using the  procedure described in \S~\ref{methanol}. We then compared the 25 GHz (5$_2$ transition) and 44~GHz fitted positions (Figure~\ref{pairs}a), and considered them a pair if:

\begin{equation}
\Delta\theta_{\rm sep} < \sqrt{\sigma_{RA_{25}}^2 + \sigma_{Dec_{25}}^2} + \sqrt{\sigma_{RA_{44}}^2 + \sigma_{Dec_{44}}^2} + 0.1\theta_{25_{\rm beam}}
\end{equation}

\noindent where $\Delta\theta_{\rm sep}$ is the angular separation, in arcseconds, between the fitted positions of the 25 GHz and 44 GHz masers. The quantity on the right is the angular separation threshold ($\sigma_{pos}$), which consists of the sum of three terms: $\sigma_{RA}$ and $\sigma_{Dec}$ are the errors on the fitted positions of the 25 and 44~GHz masers in arcseconds, and the final term is one-tenth the geometric mean of the synthesized beam of the 25~GHz data (in arcseconds). We add this additional factor to account for the extra uncertainty introduced because the two data sets were observed at different times and with different phase calibrators, so the absolute positions might differ by up to 0.1$\theta_{\rm beam}$ even when the fitted position uncertainties of the individual masers are quite small.

Of the 25~GHz \methanol\/ masers with complementary 44~GHz data, 12 of 15 have 44~GHz counterparts with $\Delta\theta_{\rm sep} <1\sigma_{pos}$; we consider these to be ``pairs'', i.e. spatially coincident.   
Figures~\ref{pairs}(a) and (b) show the distributions of maser separations and 1$\sigma_{pos}$ values for the paired masers; as shown in Figure~\ref{pairs}(a), maser pairs are co-located within $0\farcs5$.  The three remaining 25~GHz masers have 44 GHz counterparts within $\Delta\theta_{\rm sep}$ of 1.15 to $1.43\sigma_{pos}$.  These masers - G19.36-0.03\_b, G19.36-0.03\_c, and G28.28-0.36\_a - have fitted 25~GHz 5$_2$ flux densities of 91$\pm$4 mJy, 46$\pm$2 mJy, and 93$\pm$7 mJy, respectively. While on the lower end of our observed range, these flux densities are by no means exceptional. It is notable that in general, there does not appear to be a correlation between the flux density of the 5$_2$ transition and the nearest-neighbor distance. It is possible that for these two regions (G19.36$-$0.03 and G28.28$-$0.36) there is a greater absolute position mismatch than for the other targets. Thus, at the present angular resolution it is plausible that all 25~GHz masers have a detectable 44~GHz counterpart. Higher resolution observations in multiple maser transitions would be useful to further constrain the exact position coincidence and also the physical size of the maser spots.  

The 25~GHz \methanol\/ masers are weaker than their 44~GHz counterparts by a median factor of 13. The two exceptions are G22.04+0.22\_a and G35.03+0.35\_b, which are stronger than their counterparts by factors of 1.8 and 1.1, respectively. Notably, we find no correlation between the 25 GHz and 44 GHz maser flux densities or brightness temperatures for paired masers (Fig.~\ref{pairs}(c-d)), which is consistent with the results of \citet{Voronkov07}, who found no correlation between the flux densities of 25~GHz and other Class I \methanol\/ masers. Both 44~GHz and 25~GHz \methanol\/ masers have also been observed to exhibit variations in brightness on a range of timescales \citep{Sobolev2007,Pratap2007}, so the difference in observation dates between the 44~GHz and 25~GHz data may contribute to the lack of correlation in the flux densities.   Furthermore, 44~GHz masers arise from A-type \methanol, while 25~GHz masers arise from E-type \methanol; thus it is possible that this difference in parity also contributes to the lack of correlation between maser flux densities. Finally, despite the high detection rate (85\%) of 95~GHz Class~I maser emission from our sample (Table~\ref{distances}), we cannot perform a similar comparison with the 25~GHz maser results due to the absence of interferometric observations in the higher frequency line.

\subsection{Comparison with Millimeter Molecular Line Surveys of EGOs}
Here we focus on comparison with other molecular line surveys of EGOs that target complex molecules \citep[$\ge$6 atoms;][]{Herbst2009} and include a significant fraction of our sample. \citet{He2012} conducted a survey of 89 northern EGOs ($\delta>-$38$^{\circ}$) with the Arizona Radio Observatory Submillimeter Telescope (ARO SMT; beam size $\sim$29\arcsec) in multiple transitions of H$^{13}$CO$^+$, SiO, SO, CH$_3$OH, CH$_3$OCH$_3$, CH$_3$CH$_2$CN, HCOOCH$_3$, and HN$^{13}$C, c-HCCCH, and H$_2$CCO, as well as the unidentified line U260365. They detected 18 of the EGOs in our VLA sample in one or more transition (G08.67-0.35 was not targeted and G25.27-0.43 was targeted but not detected). We compared their detection rates in each line to our EGOs with and without 25~GHz thermal \methanol\/ emission to search for a correlation between the presence of 25~GHz thermal \methanol\/ and other species. We find that sources with thermal 25~GHz \methanol\/ have higher overall detection rates (considering all observed transitions) in \citet{He2012} than sources without thermal \methanol. However, this difference is primarily due to the higher detection rates that our sample have in the \citet{He2012} sample, specifically in the \methanol\/ lines. Sources with 25~GHz thermal \methanol\/ have a typical detection rate of 67-78\% in the \methanol\/ lines of the He sample, while the sources without thermal \methanol\/ have a typical detection rate of only 30-40\%. Detection rates among non-\methanol\/ species are about the same for sources with and without 25~GHz thermal \methanol. The only such species detected in the majority of our sources are H$^{13}$CO$^+$, SiO, and SO, and these detection rates are equally high for sources with and without 25~GHz thermal \methanol. Although we do not find a correlation between thermal \methanol\/ in our sample and any particular non-\methanol\/ species, we do find that sources with thermal \methanol\/ are detected in a greater {\it number} of non-\methanol\/ species than those without, indicating a possible correlation between the presence of 25~GHz thermal \methanol\/ emission and a richer gas chemistry. 

\citet{Ge2014} use the data of \citet{He2012} to determine rotational temperatures and abundances for four of the species observed (\methanol\/, CH$_3$OCH$_3$, HCOOCH$_3$, and CH$_3$CH$_2$CN). They list results for seven of the EGOs with 25~GHz thermal \methanol\/ and for four of the EGOs without. 
The EGOs with thermal \methanol\/ do not appear to be significantly hotter or cooler than those without, based on these rotational temperatures.
The median \methanol\/ abundance of the EGOs with thermal \methanol\/ is 1.43$\times10^{-9}$ and 1.06$\times10^{-9}$ for those without. While the abundance is slightly higher for sources with 25~GHz thermal \methanol\/, it is worth noting that the source with the highest abundance, G14.33-0.64, has an abundance a factor of ten higher than the median but does not have detectable 25~GHz thermal \methanol\/ emission at the current sensitivity.

\section{Conclusions}
\label{conclusions}

In a high-resolution VLA survey of 20 Extended Green Objects (EGOs) in the Milky Way, we identify 34 sites of 25~GHz Class I \methanol\/ maser emission, 10 sites of thermal \methanol\/ emission, and 20 sources of 1.3~cm continuum emission.  Thirteen of the continuum sources are new detections at 1.3~cm, having a typical peak intensity of 0.5~\mjyb.  To our knowledge, seven of these objects are new detections at any cm-$\lambda$, while 12 are either coincident with or within $2\arcsec$ of 6.7~GHz Class II \methanol\/ maser emission.  Regardless of the type of \methanol\/ emission (maser or thermal), it is strongly correlated in position with 4.5~\mum\/ EGO emission.  We also find a strong correlation between the presence of thermal \methanol\/ emission and the presence of 1.3~cm continuum emission, with the two occurring coincidentally in nine out of ten cases (see \S \ref{individual} for a discussion of the lone exception to this trend, G10.34$-$0.14). Note that the inverse relation is not true: of the twenty sources of 1.3~cm emission, only nine have coincident thermal \methanol\/ emission. 
While there is a correlation between the {\it presence} of 1.3~cm emission and 25~GHz Class I \methanol\/ masers, there is an anti-correlation between their {\it positions}. Specifically, of our 16 targets with both 1.3~cm continuum and \methanol\/ line emission, ten have 25~GHz \methanol\/ maser emission. However, only 8 of 34 masers lie within the boundaries of the 4$\sigma$ contours of the 1.3~cm continuum emission.\

For the sites classified as maser emission, the fitted flux densities are strongest in the 5$_2$ transition (see Table \ref{stats}), but the 8$_2$ transition is not significantly weaker than the 5$_2$ transition in general. The rarest transition is 10$_2$, which is detected in only 45\% of targets compared to the 60\% detection rate for 3$_2$ and the 80\% detection rates for 5$_2$ and 8$_2$.  For the 25~GHz masers for which we have complementary 44~GHz Class I \methanol\/ maser data, we find likely or possible 44~GHz companions for every 25~GHz maser, which is consistent with the suggestion that Class I \methanol\/ masers at 25~GHz and 44~GHz trace similar excitation conditions \citep{Sobolev2007}. In general, the 25~GHz masers are significantly weaker than their 44~GHz counterparts, however, we do not find any correlation between the flux densities or brightness temperatures of the paired masers. Higher matched resolution observations of masers at both wavelengths are needed in order to further constrain both the brightness temperatures and exact positions of each 25~GHz and 44~GHz maser.

\acknowledgments
The National Radio Astronomy Observatory is a facility of the National Science Foundation operated under cooperative agreement by Associated Universities, Inc. The Dunlap Institute is funded through an endowment established by the David Dunlap family and the University of Toronto. This research made use of NASA's Astrophysics Data System Bibliographic Services, APLpy (an open-source plotting package for Python hosted at http://aplpy.github.com), the SIMBAD database (operated at CDS, Strasbourg, France), and CASA.  C.~J.~Cyganowski acknowledges support from the STFC (grant number ST/M001296/1).

\end{document}